\documentclass[a4paper,11pt]{article}
\pdfoutput=1 

\usepackage{jheppub} 

\usepackage[T1]{fontenc} 

\usepackage{hyperref}
\usepackage{ulem}
\usepackage{color}
\usepackage{graphicx}
\usepackage{multirow}
\usepackage{amsmath,amssymb,amsthm,amsxtra,overpic,bbm,bm,float,epsfig}

\usepackage{color}

\newcommand{\ie}{{\it i.e.}}

\title{\boldmath Invisible neutrino decays at the MOMENT experiment}

\author[a]{Jian Tang,}
\author[a]{Tse-Chun Wang,}
\author[a,b]{and Yibing Zhang}

\affiliation[a]{School of Physics, Sun Yat-Sen University,\\ No. 135, Xingang Xi Road, Guangzhou, 510275, P. R. China}
\affiliation[b]{School of Mathematical and Physical Sciences, University of Sussex,\\ Falmer, Brighton BN1 9RH, UK}

\emailAdd{tangjian5@mail.sysu.edu.cn}
\emailAdd{wangzejun@mail.sysu.edu.cn}
\emailAdd{yz454@sussex.ac.uk}

\abstract{
We investigate invisible decays of the third neutrino mass eigenstate in future accelerator neutrino experiments using muon-decay beams such as MuOn-decay MEdium baseline NeuTrino beam experiment (MOMENT). MOMENT has outstanding potential to measure the deficit or excess in the spectra caused by neutrino decays, especially in $\nu_\mu$ and $\bar{\nu}_\mu$ disappearance channels. Such an experiment will improve the constraints of the neutrino lifetime $\tau_3$. Compared with exclusion limits in the current accelerator neutrino experiments T2K and NOvA under the stable $\nu$ assumption, we expect that MOMENT gives the bound of $\tau_3/m_3\ge10^{-11}$~s/eV at $3\sigma$, which is better than their recent limits: $\tau_3/m_3\ge 7\times10^{-13}$~s/eV in NOvA and $\tau_3/m_3\ge 1.41\times10^{-12}$~s/eV in T2K.
The non-decay scenario is expected to be excluded by MOMENT at a confidence level $>3\sigma$, if the best fit results in T2K and NOvA are confirmed.
We further find that reducing systematic uncertainties is more important than the running time.
Finally, we find some impact of $\tau_3/m_3$ on the precision measurement of other oscillation parameters.
}

\begin{document} 
\maketitle
\flushbottom

\section{Introduction}
\label{intro}
The oscillation pattern of three-flavour neutrino mixing has been established through solar, atmospheric, accelerator and reactor neutrino
experiments~\cite{Aharmim:2011vm,Wendell:2010md,Abe:2008aa,An:2016ses}. In the standard three-flavour 
paradigm, neutrino oscillations are dominated by two mass-squared splittings (\ie, $\Delta m^2_{31}$, $\Delta m^2_{21}$) and three mixing angles
(\ie~$\theta_{12}$, $\theta_{13}$, $\theta_{23}$)~\cite{Patrignani:2016xqp}.
Up to now, most of the oscillation parameters have been measured well~\cite{Esteban:2016qun}, except the Dirac CP phase $\delta$ and the neutrino mass ordering (normal mass hierarchy: $\Delta m^2_{31}>0$; inverted mass hierarchy: $\Delta m^2_{31}<0$).
The precision of measuring $\theta_{23}$ is not good enough to discriminate the octant degeneracies with a specific prediction $\theta_{23}=45^{\circ}$. All these unknown parameters will be measured
in the near future by medium baseline reactor experiments: JUNO~\cite{Ranucci:2017cek} and RENO~\cite{Seo:2015yqp}, and by the long-baseline accelerator neutrino 
experiments: T2K ~\cite{Abe:2014tzr}, NOvA~\cite{Adamson:2016xxw}, T2HK~\cite{Abe:2014tzr} and DUNE ~\cite{Acciarri:2015uup}.
Recent results from T2K and NOvA incline to a normal mass hierarchy and indicate a hint of $\delta\approx
270^\circ$~\cite{Abe:2017uxa,Adamson:2017gxd} only at a low confidence level. Therefore, we are looking forward to data provided by the next-generation experiments to attain a compelling conclusion. Since we are entering an era of precision measurements, it is natural to expect near future neutrino oscillation experiments to search for new physics beyond three-generation neutrino oscillations including sterile neutrinos, neutrino decays and non-standard neutrino interactions, and so on.

Neutrino decays are classified into invisible and visible scenarios.
%
%
Several models depend on whether neutrinos are Majorana or Dirac particles~\cite{Acker:1991ej,Acker:1993sz,Gelmini:1980re,Chikashige:1980ui,Pakvasa:1999ta,Kim:1990km,Acker:1992eh,Lindner:2001fx}. 
If the final states of neutrino decays are unobservable to the detector, those decays are called invisible decays~\cite{Lindner:2001fx}. There are decay models
$\nu_j\rightarrow\nu_4+J$ for Majorana neutrinos~\cite{Pakvasa:1999ta,Chikashige:1980ui,Gelmini:1980re}, where $J$ denotes a Majoron. Another class of models assumes that neutrinos are Dirac particles and the coupling which gives rise to neutrino decay: $\nu_i\rightarrow\bar{\nu}_{jR}+\chi$, where $\chi$ is a light iso-singlet scalar and $\nu_{iR}$ is a right-handed fermion~\cite{Acker:1991ej,Acker:1993sz}. 
In the visible decay scenario, decay products can be detected by the detector. Several decay patterns like $\nu_j\rightarrow \bar{\nu}_i(\nu_i)+J$ have been put forward\cite{Kim:1990km,Acker:1992eh,Lindner:2001fx}. 

The $\nu_2$ decay in the invisible channel has been constrained well from solar neutrino oscillation data, which gives the bound $\tau_2/m_2>7.2\times10^{-4}$~s/eV at $90\%$ C.L.~\cite{Picoreti:2015ika,Berryman:2014qha}.  There are proposals to constrain the neutrino decays life time with the help of solar neutrino oscillations detected by the liquid Xenon detector~\cite{Huang:2018nxj}.
Atmospheric and long-baseline neutrino experiments set a bound on the decay lifetime for $\nu_3$, such as $\tau_3/m_3>2.9\times10^{-10}$~s/eV at $90\%$ C.L.~\cite{GonzalezGarcia:2008ru}. Recently, invisible neutrino decays have been used to explain the IceCube track and cascade tension~\cite{Denton:2018aml}. A sensitivity study of invisible neutrino decays has been conducted for KM3NeT-ORCA~\cite{deSalas:2018kri}.  
MINOS and T2K experiments have constrained the neutrino decay lifetime as $\tau_3/m_3>2.8\times10^{-12}$~s/eV at $90\%$ confidence level~\cite{Gomes:2014yua}.
\textbf{Recently a combined analysis of NOvA and T2K data points to a result of} $\mathbf{\tau_3/m_3>1.5\times10^{-12}}$ \textbf{s/eV along with the constraints by individual experiments:} $\mathbf{\tau_3/m_3\ge 7\times10^{-13}}$\textbf{~s/eV in NOvA and} $\mathbf{\tau_3/m_3\ge 1.41\times10^{-12}}$\textbf{~s/eV in T2K~\cite{Choubey:2018cfz}.}
The expected bounds for JUNO~\cite{Abrahao:2015rba}, INO\cite{Choubey:2017eyg} and DUNE~\cite{Choubey:2017dyu} can reach
$\tau_3/m_3>7.5$ $(5.5)\times10^{-11}$~s/eV at $95\%$ (99$\%$) C.L., $\tau_3/m_3>1.51\times 10^{-10}$~s/eV at $90\%$ C.L. and $\tau_3/m_3>4.5\times10^{-11}$~s/eV at $90\%$ C.L., respectively. 

Some studies focus on visible decays. For example, a study shows that DUNE will be sensitive to the level of $\tau_3/m_3<1.95-2.6\times10^{-10}$ s/eV at $>90\%$ C.L., and the combination of MINOS and T2K gives the bound $\tau_3/m_3>1.5\times10^{-11}$ s/eV at a confidence level $>90\%$~\cite{Gago:2017zzy}. As visible neutrino decays offer clear signals in the detector, it is even more difficult to constrain invisible decays than in the visible case. Because the bound for the invisible neutrino decay like $\tau_3$ is much worse than $\tau_2$, it is valuable to exploit the measurement potential of invisible $\nu_3$ decays in next generation neutrino oscillation experiments.
We further point out that in addition to searching for them in neutrino oscillation experiments, we can also find evidence for neutrino decays in astrophysical observations due to their influence on the formation of cosmological perturbations~\cite{Khlopov1,Khlopov2}.

Apart from superbeam neutrino experiments, it is desirable to study new physics at muon-decay accelerator neutrino experiments. 
%
%
In such experiments, neutrinos come from a three-body decay process, avoiding intrinsic electron-flavor neutrino contaminations in the reconstructed oscillation signals from the source. 
Apart from such an advantage, MOMENT~\cite{Cao:2014bea} is likely to use a Gd-doped water cherenkov detector capable of detecting multiple channels, which have been demonstrated to have excellent properties to study new physics, including NSIs~\cite{Gavela:2008ra,Bonnet:2009ej,Krauss:2011ur} 
and sterile neutrinos~\cite{Gariazzo:2017fdh,Abazajian:2012ys,Adhikari:2016bei,Minkowski:1977sc}. 
In the current work, we focus on the constraints of neutrino decays
into invisible products, and demonstrate how the $\nu_3$ decay would affect precision measurements of standard neutrino mixing parameters. 

This paper is organized as follows: we describe the basic framework for neutrino oscillations with invisible neutrino decays taken into account and study the oscillation probabilities for the MOMENT experiment in Sec.~\ref{sec:prob}. Implementations and simulation details are given in Sec.~\ref{sec:simulation}, and in the same section, we also investigate the impact of neutrino decays on the spectra of MOMENT. In Sec.~\ref{sec:results}, we present simulation results, mainly focusing on the constraints on the $\nu_3$ lifetime, compare it to the reach of current experiments, and investigate the impacts of the total running time, systematic uncertainty and energy resolution on this measurement, with the study on the expected exclusion level to the stable-neutrino assumption and their impacts on precision measurements of $\theta_{23}$ and $\Delta m^2_{31}$. Finally, we summarize in Sec.~\ref{sec:summary}.

\section{Neutrino oscillations with invisible neutrino decays}
\label{sec:prob}

The latest results from MiniBooNE have an excess for reconstructed oscillation spectra~\cite{Aguilar-Arevalo:2018gpe}, suggesting the existence of sterile neutrinos. We assume that the neutrino decay products are sterile neutrinos. 
In addition, we consider that the third mass eigenstate decays in the following channel: $\nu_3\rightarrow\nu_4+J$, where normal mass hierarchy and a light sterile neutrino are considered (\ie~$m_3 > m_2 > m_1> m_4$). 
The connection between flavour eigenstates and mass eigenstates can be given as: 
\begin{equation}
\left(
  \begin{array}{cc}
  \nu_{\alpha}\\
   \nu_s\\
  \end {array}
\right)
=
\left(
  \begin{array}{cc}
  U&0\\
  0&1\\
  \end {array}
\right)
\left(
  \begin{array}{cc}
  \nu_i\\
  \nu_4\\
  \end {array}
\right)
\label{decay_eq1}
\end{equation}
The Hamitonian of neutrino propagation in matter can be written as:
\begin{eqnarray}
H=
  U\left\{
  \frac{1}{2E}\left(
  \begin{array}{ccc}
  0 & 0 & 0\\
  0 & \Delta m_{21}^2 & 0\\
  0 & 0 &\Delta m_{31}^2\\
  \end {array}
  \right)-i\frac{m_3}{2E\tau_3}
   \left(
    \begin{array}{ccc}
  0 & 0 &0\\
  0 & 0 &0\\
  0 & 0 &1\\
  \end {array}
  \right)
  \right\}U^{\dag}
  +
  \left(
    \begin{array}{ccc}
  2\sqrt{2}G_FN_eE & 0 &0\\
  0 & 0 &0\\
  0 & 0 &0\\
  \end {array}
  \right)
  \label{eq:hamiton_decay},
\end{eqnarray}
where $U$ is the PMNS mixing matrix~\cite{Maki:1962mu,Pontecorvo:1957cp}, $G_F$ is the Fermi coupling constant, $N_e$ is the electron density, $E$ is the neutrino energy and $\tau_3$ is the lifetime of $\nu_3$. Obviously, the probabilities for neutrino and antineutrino modes remain invariant with a replacement of $\delta\rightarrow -\delta$ and $N_e\rightarrow -N_e$, \ie~$P_{\nu_\alpha\rightarrow\nu_\beta}(E,L;\delta,N_e)=P_{
\bar{\nu}_\alpha\rightarrow\bar{\nu}_\beta}(E,L;-\delta,-N_e)$.
Then we can calculate the numerical oscillation probabilities by diagonalizing the Hamitonian matrix. The diagonalization method can be found in Ref.~\cite{Hahn:2006hr}. Our numerical tool to evaluate the probabilities with neutrino decays has been checked by comparing our result with those shown in Ref.~\cite{Choubey:2017dyu}.
To cross check validity of our codes, we have reproduced the invisible-neutrino-decay result from Ref.~\cite{Choubey:2018cfz}, highlighting the current measurement at T2K and NOvA.
The probability for the antineutrino mode has been cross checked by a comparison with the neutrino mode taking the opposite sign of $\delta$ and $N_e$.


\begin{figure}[!t]%
\centering
\includegraphics[width=2.5in]{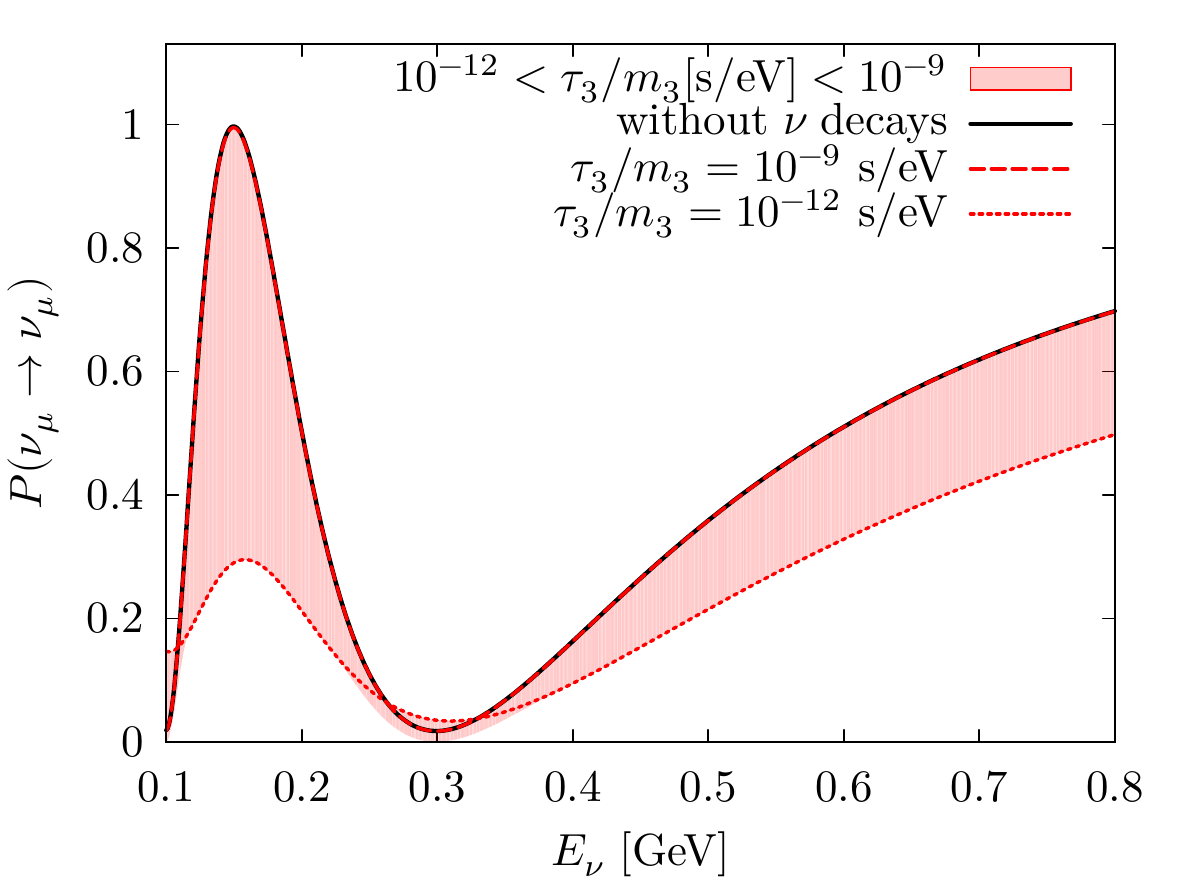}
\includegraphics[width=2.5in]{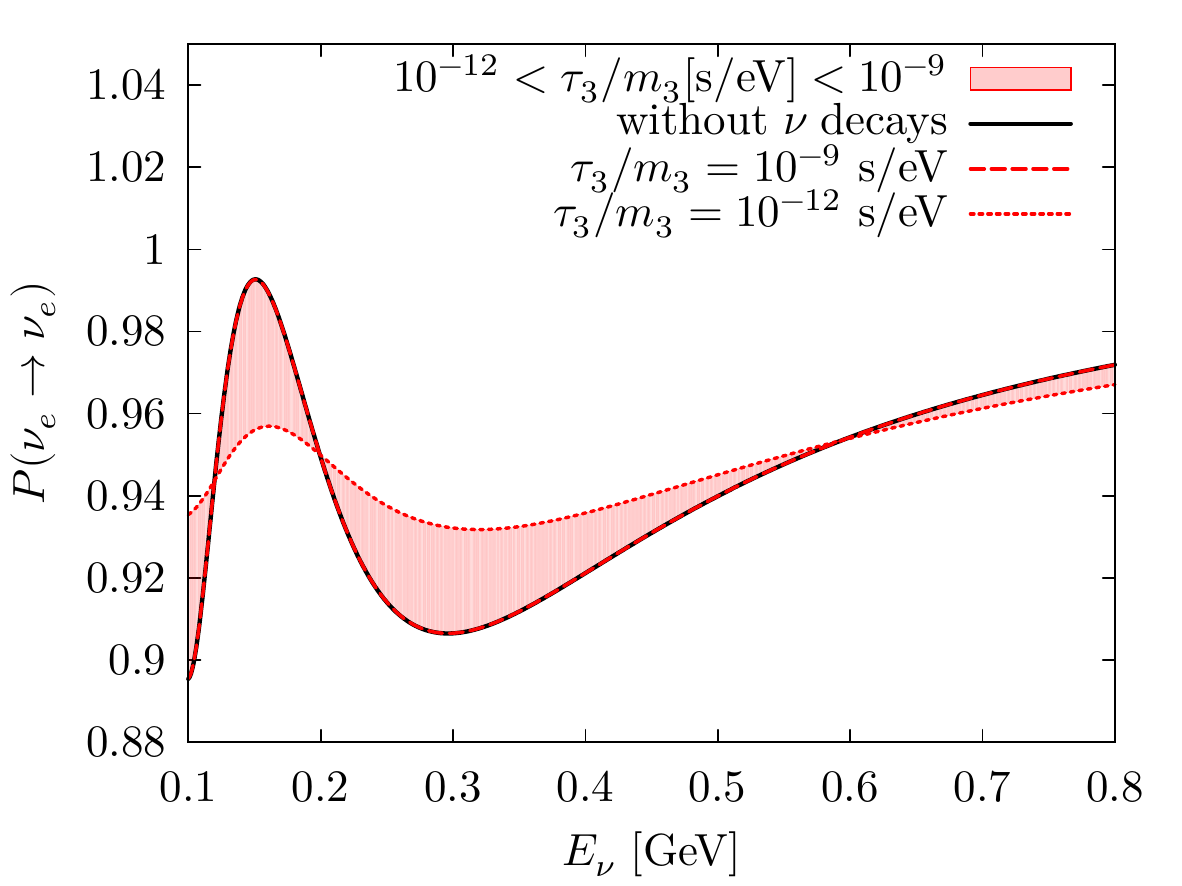}\\
\includegraphics[width=2.5in]{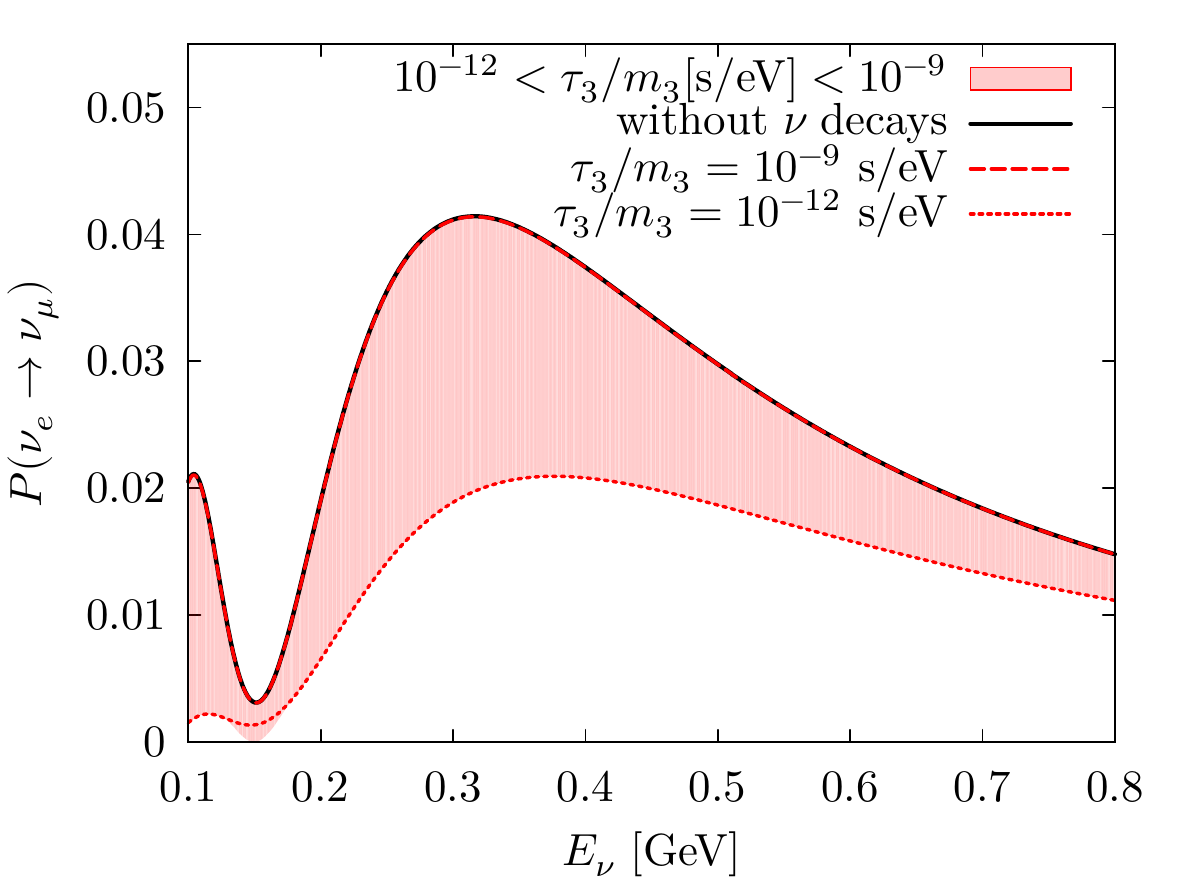}
\includegraphics[width=2.5in]{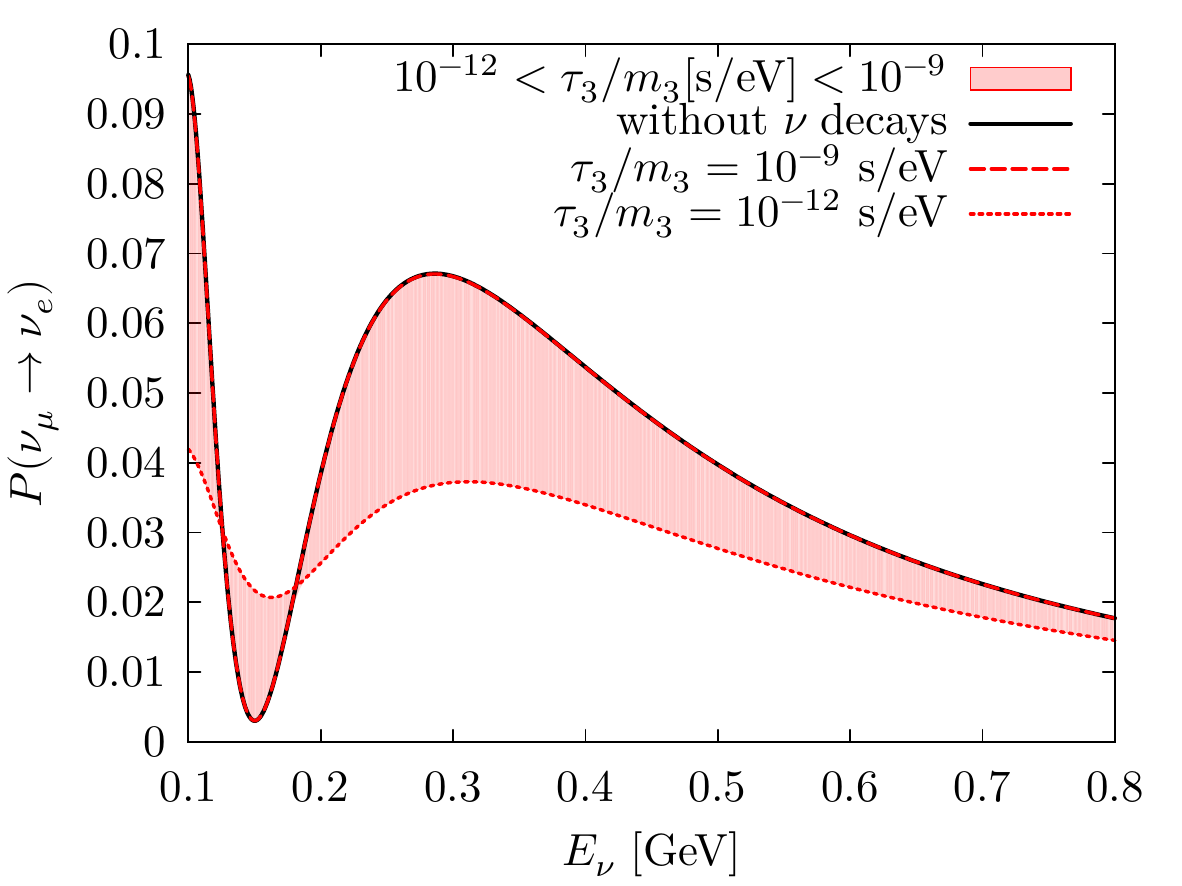}
\caption{The oscillation probabilities within $10^{-12}$ s/eV $<\tau_3/m_3<10^{-9}$ s/eV (red band) for MOMENT. We especially present the probability with $\tau_3/m_3=10^{-12}$ s/eV (red-dotted curve), $=10^{-9}$ s/eV (red-dashed curve) and $=\infty$ (black curve). Four channels are shown: $\nu_\mu\rightarrow\nu_\mu$ (upper-left), $\nu_e\rightarrow\nu_e$ (upper-right), $\nu_e\rightarrow\nu_\mu$ (right-left), and $\nu_\mu\rightarrow\nu_e$ (lower-right). The following oscillation parameters are used: $\theta_{12}$=$33.8^{\circ}$,
$\theta_{13}$=$8.61^{\circ}$, $\theta_{23}$=$49.6^{\circ}$, $\Delta m_{21}^2$=7.39$\times10^{-5}$~eV$^2$, $\Delta m_{31}^2$=2.52$\times10^{-3}$~eV$^2$, and $\delta =270^\circ$.}%
\label{fig:prob}
\end{figure}

\begin{figure}[!h]%
\centering
\includegraphics[width=2.5in]{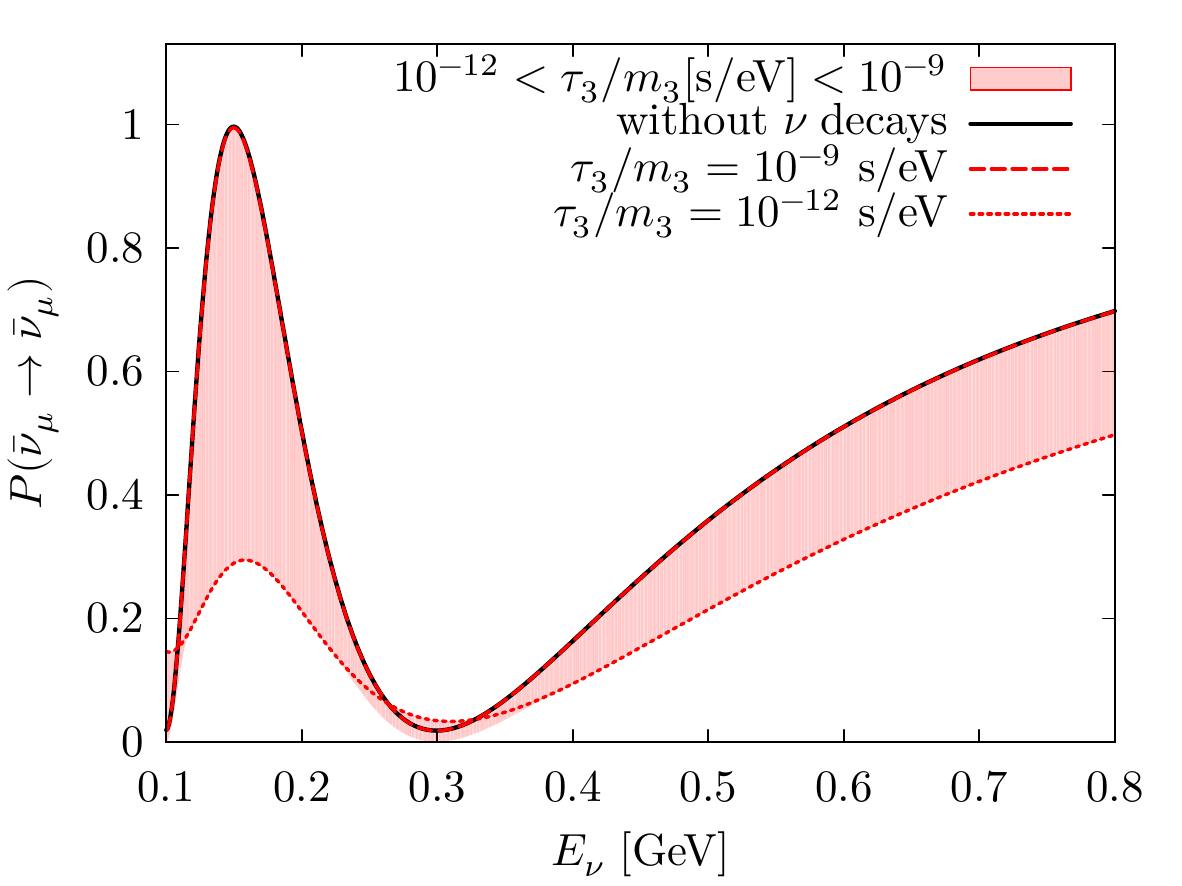}
\includegraphics[width=2.5in]{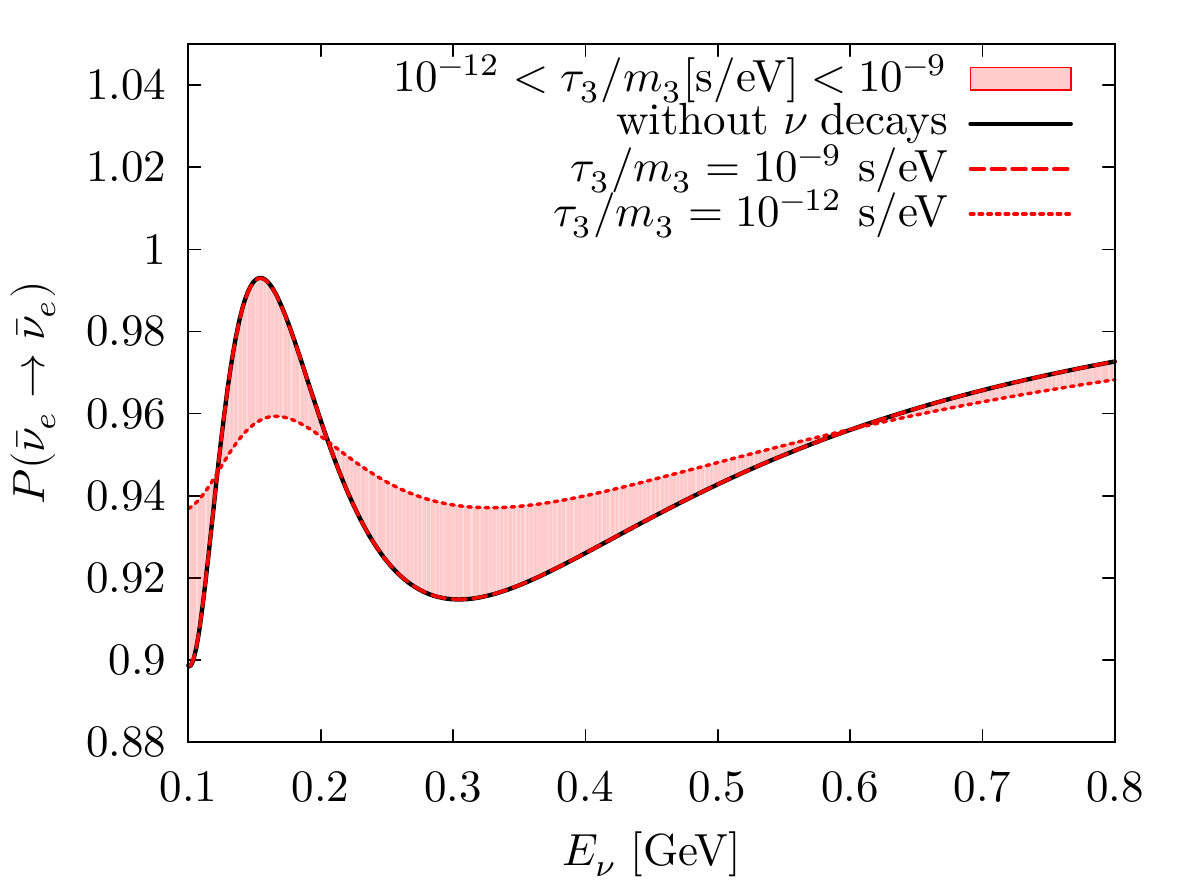}\\
\includegraphics[width=2.5in]{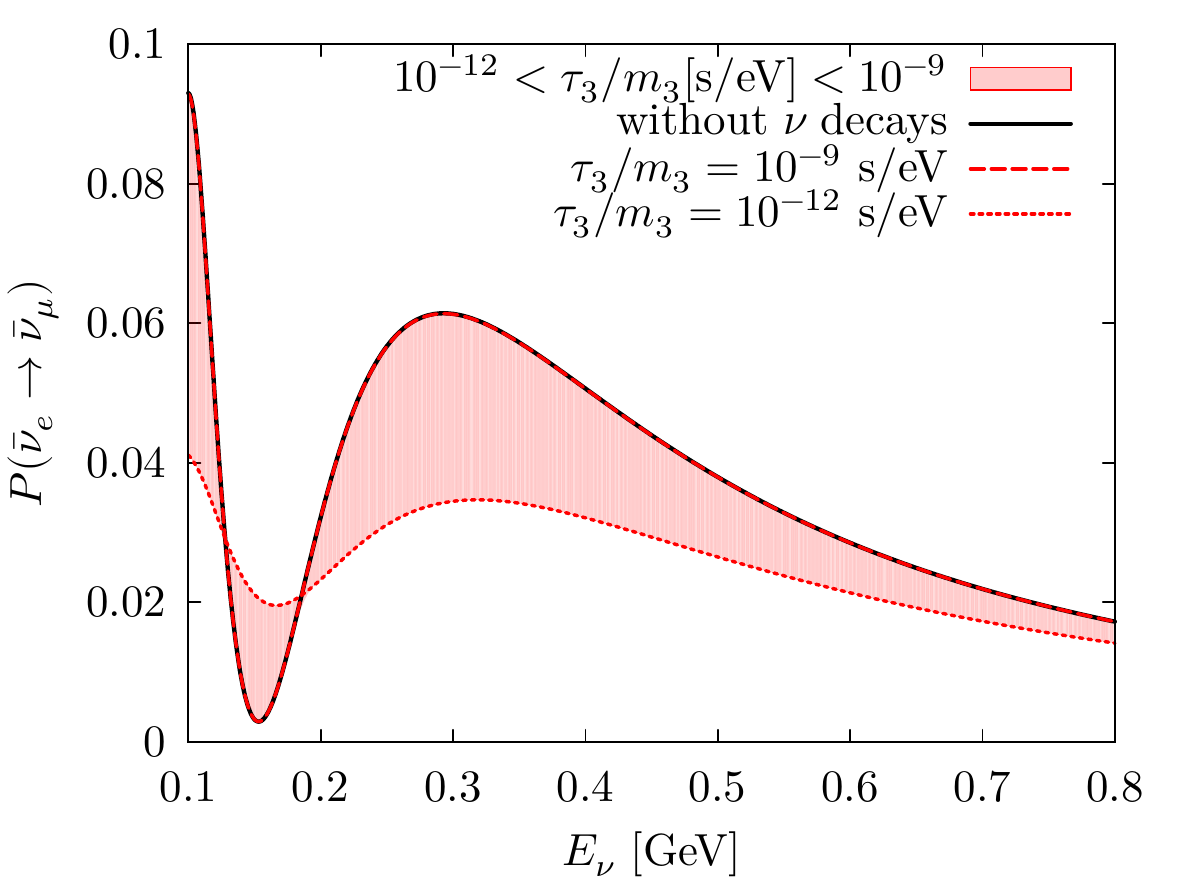}
\includegraphics[width=2.5in]{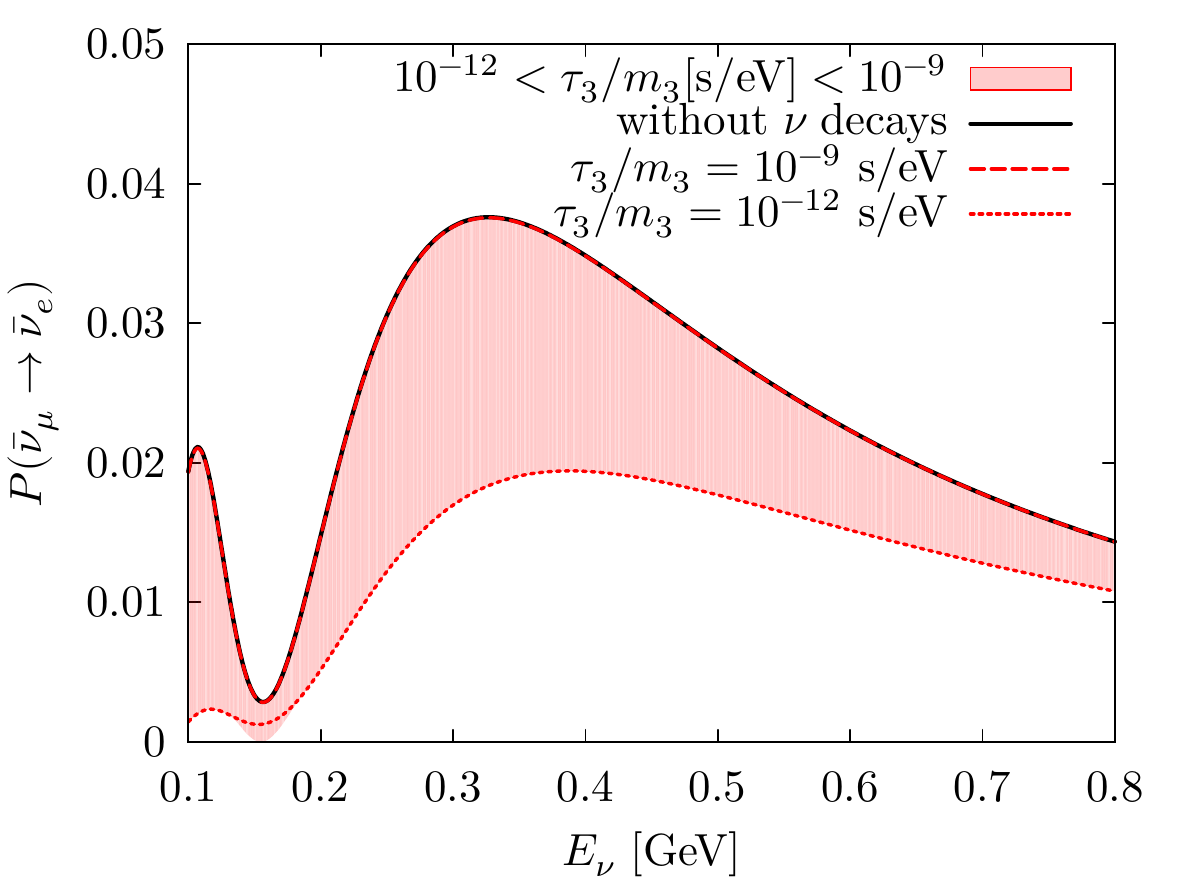}
\caption{The oscillation probabilities within $10^{-12}$ s/eV $<\tau_3/m_3<10^{-9}$ s/eV (red band) for MOMENT. We especially present the probability with $\tau_3/m_3=10^{-12}$ s/eV (red-dotted curve), $=10^{-9}$ s/eV (red-dashed curve) and $=\infty$ (black curve). Four channels are considered: $\bar{\nu}_\mu\rightarrow\bar{\nu}_\mu$ (upper-left), $\bar{\nu}_e\rightarrow\bar{\nu}_e$ (upper-right), $\bar{\nu}_e\rightarrow\bar{\nu}_\mu$ (right-left), and $\bar{\nu}_\mu\rightarrow\bar{\nu}_e$ (lower-right). The following oscillation parameters are used: $\theta_{12}$=$33.8^{\circ}$, $\theta_{13}$=$8.61^{\circ}$, $\theta_{23}$=$49.6^{\circ}$, $\Delta m_{21}^2$=7.39$\times10^{-5}$~eV$^2$, $\Delta m_{31}^2$=2.52$\times10^{-3}$~eV$^2$, and $\delta =270^\circ$.}%
\label{fig:prob_bar}
\end{figure}

The probabilities with $\nu$ decays in vacuum are given as follows,
\begin{equation}
\begin{aligned}
P_{\nu_\alpha\rightarrow\nu_\beta}(E, L; \delta)
=\left|
U_{\alpha 1}(\delta)U^*_{\beta 1}(\delta) +U_{\alpha 2}(\delta)U^*_{\beta 2}(\delta) \exp\left(-i \frac{\Delta m^2_{21}}{2E}\right)\right.\\
\left.+U_{\alpha 3}(\delta)U^*_{\beta 3}(\delta) \exp\left(-\Gamma_3L\right)\exp\left(-i \frac{\Delta m^2_{31}}{2E}\right)\right|^2.
\end{aligned}
\end{equation}
This can be further expanded as 

\begin{equation}\label{eq:prob_expansion}
\begin{array}{l}
P_{\nu_\alpha\rightarrow\nu_\beta}(E, L; \delta)\\
=U^*_{\alpha 1}(\delta)U_{\beta 1}(\delta) U_{\alpha 1}(\delta)U^*_{\beta 1}(\delta) + U^*_{\alpha 2}(\delta)U_{\beta 2}(\delta) U_{\alpha 2}(\delta)U^*_{\beta 2}(\delta)\\
+ U^*_{\alpha 3}(\delta)U_{\beta 3}(\delta) U_{\alpha 3}(\delta)U^*_{\beta 3}(\delta)\exp(-2\Gamma_3 L)
\\+\text{Re}\left[ U^*_{\alpha 2}(\delta)U_{\beta 2}(\delta) U_{\alpha 1}(\delta)U^*_{\beta 1}(\delta)  \right]
\cos\left(\frac{\Delta m_{21}^2 L}{2E}\right) \\
+\text{Im}\left[ U^*_{\alpha 2}(\delta)U_{\beta 2}(\delta) U_{\alpha 1}(\delta)U^*_{\beta 1}(\delta)  \right]\sin\left(\frac{\Delta m_{21}^2 L}{2E}\right)
\\+\text{Re}\left[ U^*_{\alpha 3}(\delta)U_{\beta 3}(\delta) U_{\alpha 1}(\delta)U^*_{\beta 1}(\delta)  \right]
\exp(-\Gamma_3 L)\cos\left(\frac{\Delta m_{31}^2 L}{2E}\right) 
\\+\text{Im}\left[ U^*_{\alpha 3}(\delta)U_{\beta 3}(\delta) U_{\alpha 1}(\delta)U^*_{\beta 1}(\delta)  \right]\exp(-\Gamma_3 L)\sin\left(\frac{\Delta m_{31}^2 L}{2E}\right)
\\+\text{Re}\left[ U^*_{\alpha 3}(\delta)U_{\beta 3}(\delta) U_{\alpha 2}(\delta)U^*_{\beta 2}(\delta)  \right]
\exp(-\Gamma_3 L)\cos\left(\frac{\Delta m_{32}^2 L}{2E}\right) 
\\+\text{Im}\left[ U^*_{\alpha 3}(\delta)U_{\beta 3}(\delta) U_{\alpha 2}(\delta)U^*_{\beta 2}(\delta)  \right]\exp(-\Gamma_3 L)\sin\left(\frac{\Delta m_{32}^2 L}{2E}\right),
\end{array}
\end{equation}
where $\Gamma_3\equiv\frac{m_3}{2E\tau_3}$. For the antineutrino mode, $\delta$ is replaced by $-\delta$. 
Eq.~(\ref{eq:prob_expansion}) is consistent with Eq.~(A.2) in \cite{Abrahao:2015rba}.
It is clear that through the final 4 terms of Eq.~(\ref{eq:prob_expansion}) neutrino decays provide damping effects to the $\Delta m^2_{31}$ and $\Delta m^2_{32}$ oscillations. Further, the decays also cause an overall decrease via the third term. Both effects can be seen in the following.

We show the probability for four channels $\nu_\mu\rightarrow\nu_\mu$ (upper-left), $\nu_e\rightarrow\nu_e$ (upper-right), $\nu_e\rightarrow\nu_\mu$ (lower-left), and $\nu_\mu\rightarrow\nu_e$ (lower-right) of MOMENT in Fig.~\ref{fig:prob} (those for the antineutrino mode are in Fig.~\ref{fig:prob_bar}). For the case with neutrino decays, we consider those within $10^{-12}$ s/eV $<\tau_3/m_3<10^{-9}$ s/eV (red band), and compare it with that for the case without neutrino decays (black curve). As we can see, the case with $\tau_3/m_3=10^{-9}$ s/eV is overlapping with the curves for the case without neutrino decays. For the other extreme case $\tau_3/m_3=10^{-12}$ s/eV, the probabilities are far from the black curves. In the following, we compare the case for $\tau_3/m_3=10^{-12}$ s/eV and that without neutrino decays.
Except for the minima, in the $\nu_\mu\rightarrow\nu_\mu$ channel, we see significant deficits. Around the minima, we notice the fact that the probability with neutrino decays goes above or below the curve corresponding to the stable-neutrino assumption. 
This is because the suppression term dominates the damping ones.
Moving to the smaller $\tau_3/m_3$, $U^*_{\alpha 3}(\delta)U_{\beta 3}(\delta) U_{\alpha 3}(\delta)U^*_{\beta 3}(\delta)\exp(-2\Gamma_3 L)$ gets smaller earlier than the damping terms because of the factor of $2$ in the exponential. When this effect does not dominate the damping one, the probability goes upper around the minima. The competition between these two effects is also seen in the $\nu_e\rightarrow\nu_\mu$ and $\bar{\nu}_\mu\rightarrow\bar{\nu}_e$ channels.
%
Therefore, the maxima in the $\nu_\mu$ and $\bar{\nu}_\mu$ disappearance channels could be useful for measuring the effect of neutrino decays.
The damping effect in $\nu_e$ and $\bar{\nu}_e$ disappearance channels is obvious.
Further, we see an overall decrease in  $P(\nu_e\rightarrow\nu_\mu)$, while the impact of neutrino decays on $P(\nu_\mu\rightarrow\nu_e)$ is similar to that for $e$ disappearance channels --- it smoothens out the probability (damping effects). The amount of impact in $P(\nu_\mu\rightarrow\nu_e)$ is similar to that in $P(\nu_e\rightarrow\nu_\mu)$. 
We see similar results for the antineutrino mode, except for the opposite pattern in the appearance channels: $P(\nu_\mu\rightarrow\nu_e)\sim P(\bar{\nu}_e\rightarrow\bar{\nu}_\mu)$ and $P(\nu_e\rightarrow\nu_\mu)\sim P(\bar{\nu}_\mu\rightarrow\bar{\nu}_e)$.
Based on the size of variations, we reach the conclusion that the $\mu$-flavour disappearance channel is the more important  than the other channels in the measurement of $\tau_3/m_3$.

\section{Simulated spectra with neutrino decays in MOMENT}
\label{sec:simulation}

The simulation details for MOMENT are shown in Table~\ref{tab:glbtable} with the neutrino sources, detector descriptions and running time~\cite{Tang:2017qen,Tang:2017khg}. MOMENT, as a medium muon decay accelerator neutrino experiment, is proposed as a future experiment to measure the leptonic CP-violating phase. The neutrino fluxes are kindly offered by the MOMENT working group~\cite{Cao:2014bea}. Here we utilize eight oscillation channels: $\nu_e\rightarrow \nu_e$, $\nu_e\rightarrow \nu_{\mu}$, $\nu_{\mu} \rightarrow \nu_e$, $\nu_{\mu} \rightarrow \nu_{\mu}$ and their CP-conjugate partners. We have to consider flavour and charge identifications to distinguish secondary particles by means of an advanced neutrino detector. The charged-current 
interactions are used to identify neutrino signals: $\nu_e + n \rightarrow p + e^-$, $\bar{\nu}_{\mu} + p \rightarrow n + \mu^+$, $\bar{\nu}_e + p \rightarrow n + e^+$, and $\nu_{\mu} + n \rightarrow p + \mu^- $. We consider the new technology using Gd-doped water to separate both Cherenkov and coincident signals from the capture of thermal neutrons~\cite{Campagne:2006yx,Ishida:2013kba}. 
The major backgrounds are mostly from the atmospheric neutrinos, neutral current backgrounds and charge mis-identifications. They can be largely suppressed by the beam direction and proper modelling of background spectra within the beam-off period, which is to be extensively studied in detector simulations. In Sec.~\ref{sec:exp_impacts}, we will compare the physics capabilities under different assumptions, including a change of total running time.

\begin{table}[h]
\centering
\begin{tabular}{cc}
\hline\hline
Experiments & MOMENT \\ \hline
Fiducial mass\hphantom{00} & \hphantom{0}Gd-doped Water cherenkov(500 kton)\\ \hline
Channels\hphantom{00} & \hphantom{0}{$\nu_e(\bar{\nu}_e)\rightarrow\nu_e(\bar{\nu}_e)$, $\nu_{\mu}(\bar{\nu}_{\mu})\rightarrow\nu_{\mu}(\bar{\nu}_{\mu})$,}  \\ 
&$\nu_e(\bar{\nu}_e)\rightarrow\nu_{\mu}(\bar{\nu}_{\mu})$, 
$\nu_{\mu}(\bar{\nu}_{\mu})\rightarrow\nu_e(\bar{\nu}_{e})$\\ \hline
Energy resolution\hphantom{0} & $12\%/E$ \\ \hline
Runtime & $\mu^-$ mode 5 yrs+ $\mu^+$ mode 5 yrs \\ \hline
Baseline & 150 km \\ \hline
Energy range  & 100 MeV to 800 MeV \\ \hline
Normalization & appearance channels: $2.5\%$ \\
(error on signal) & disappearance channels: 5$\%$\\ \hline
Normalization & {Neutral current, Atmospheric neutrinos}\\
(error on background) & Charge misidentification\\ \hline\hline
\end{tabular}
\caption{Assumptions for the source, detector and running time for MOMENT in the simulation.}
\label{tab:glbtable}
\end{table}

Our simulation is carried out with the help of a GLoBES package~\cite{Huber:2004ka,Huber:2007ji}. The following central values and their uncertainties of the standard neutrino oscillation parameters are taken from the latest NuFit4.0 results~\cite{Esteban:2016qun}: 
$\theta_{12}$=$33.82^{\circ}$ ($2.3\%$),
$\theta_{13}$=$8.61^{\circ}$ ($1.8\%$),
$\theta_{23}$=$49.6^{\circ}$ ($5.8\%$),
$\Delta m_{21}^2$=7.39$\times10^{-5}$~eV$^2$ ($2.4\%$),
$\Delta m_{31}^2$=2.525$\times10^{-3}$~eV$^2$ ($1.6\%$),
$\delta =270^\circ$ (no prior applied).
In the following, we will assume the normal mass hierarchy, \ie~$\Delta m_{31}^2>0$. 

\begin{figure}[!h]%
 \includegraphics[width=3in]{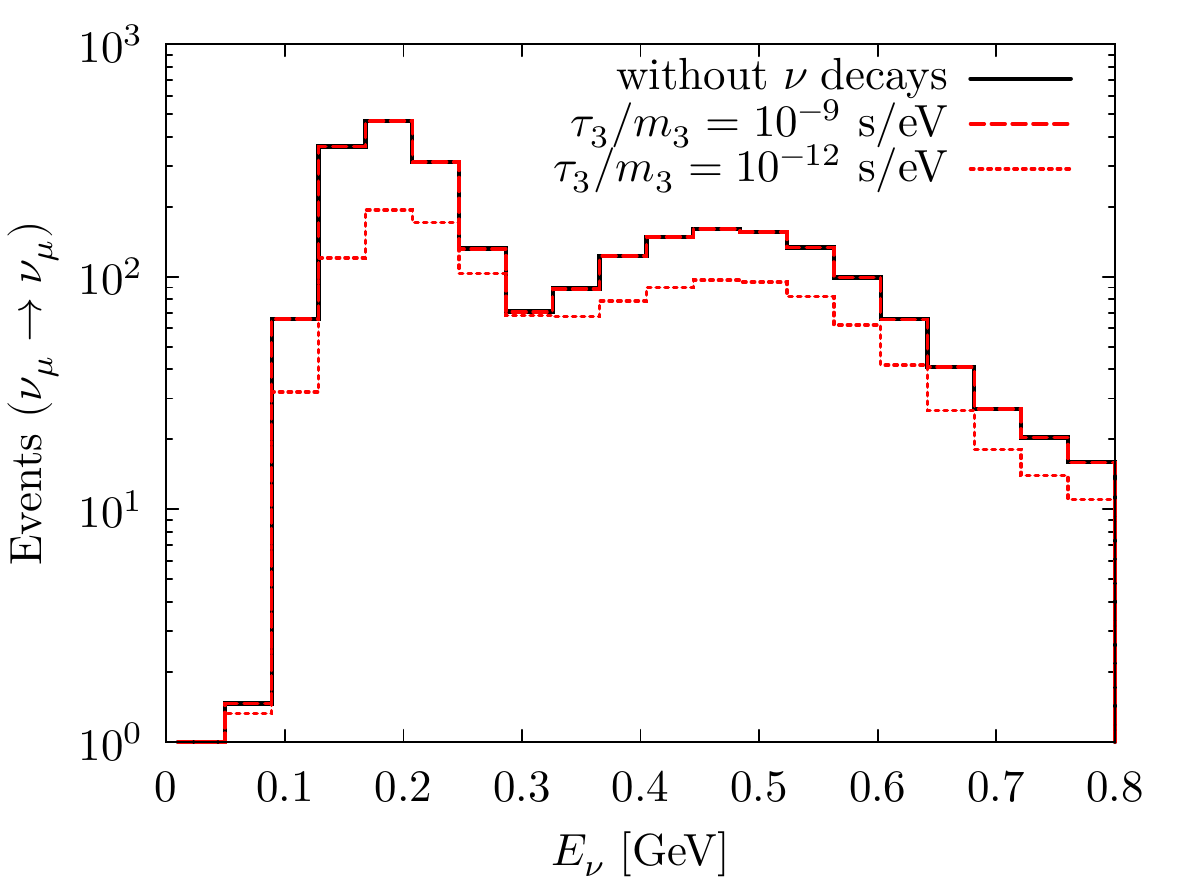}
  \includegraphics[width=3in]{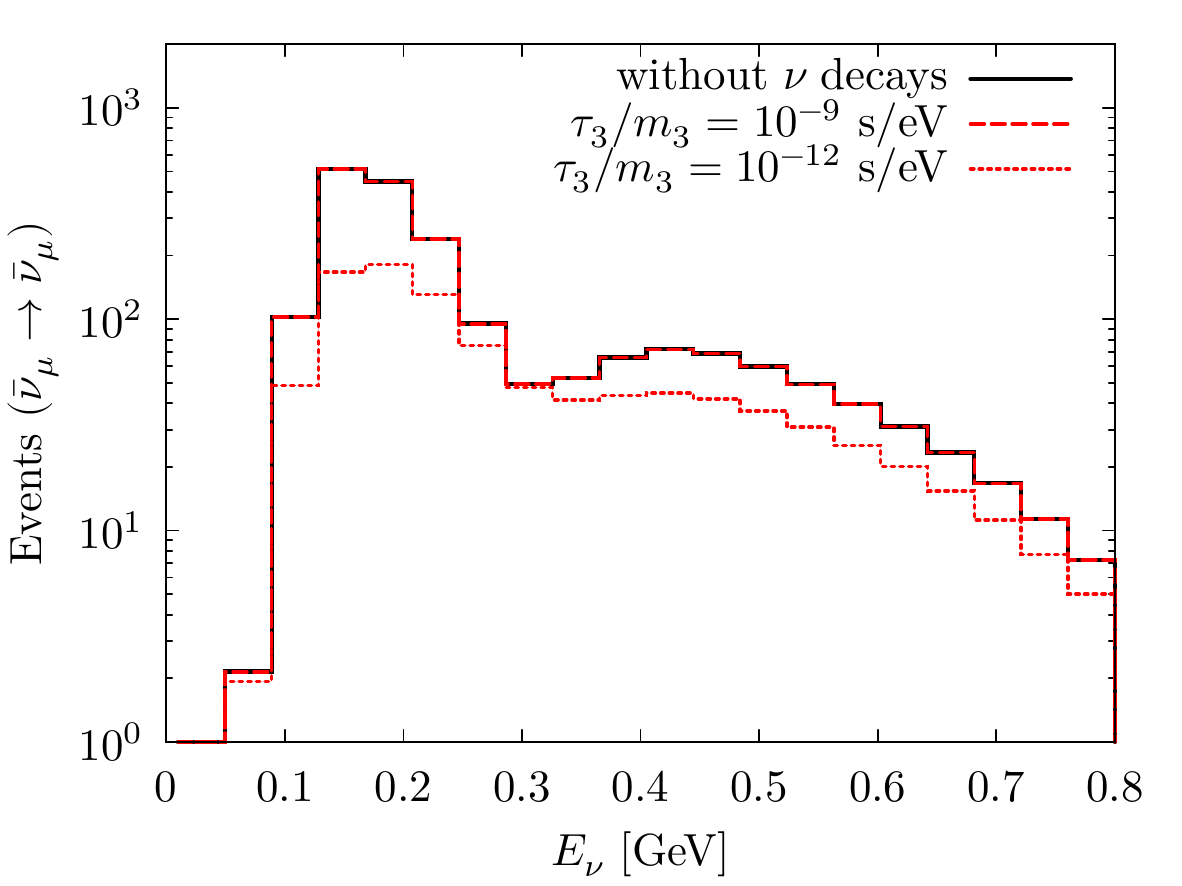}
 \includegraphics[width=3in]{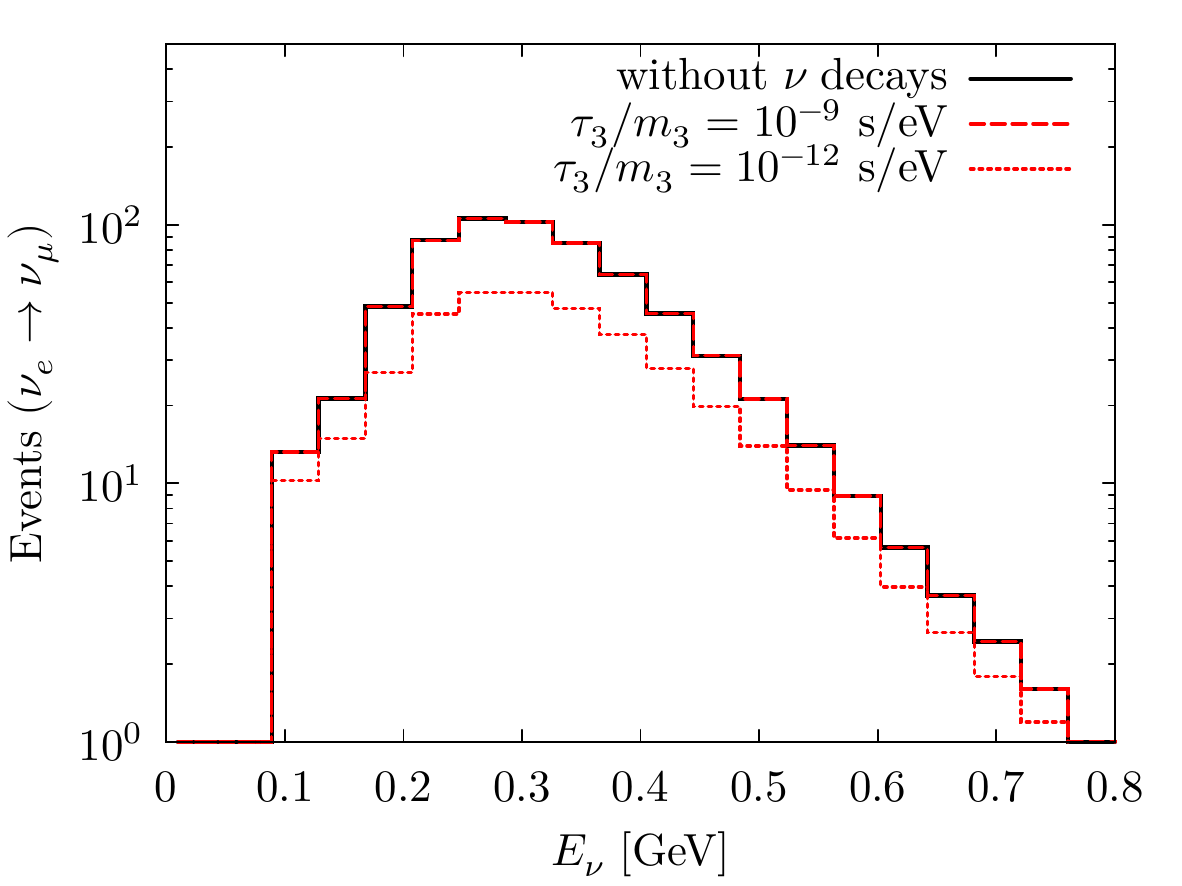}
  \includegraphics[width=3.in]{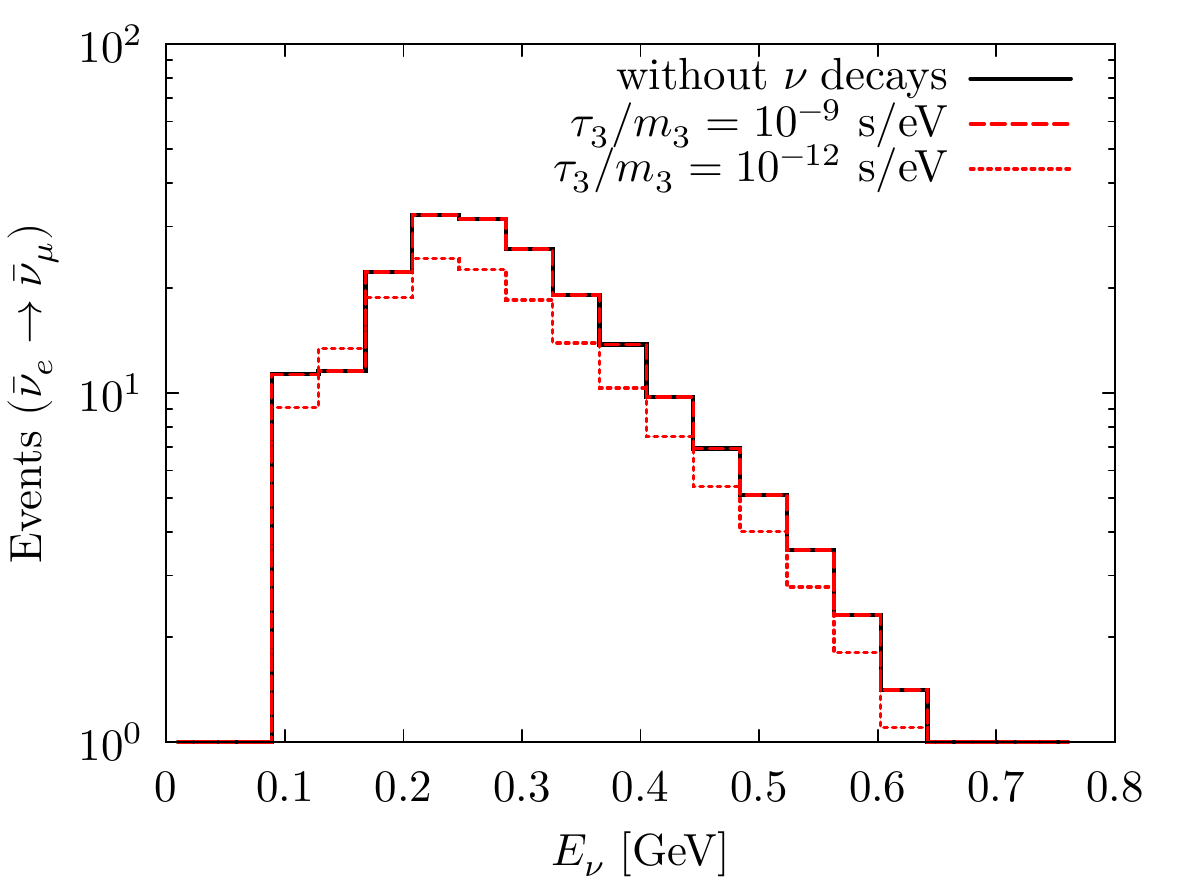}
 \caption{The spectra with $\tau_3/m_3=10^{-12}$~s/eV (red-dotted) and $=10^{-9}$~s/eV (red-dashed) and under the standard model (black) for MOMENT. Four channels are considered: $\nu_\mu\rightarrow\nu_\mu$ (upper-left), $\bar{\nu}_\mu\rightarrow\bar{\nu}_\mu$ (upper-right), $\nu_e\rightarrow\nu_\mu$ (right-left), and $\bar{\nu}_e\rightarrow\bar{\nu}_\mu$ (lower-right). The oscillation baseline is set at $150$ km.}%
 \label{fig:event_mu}
\end{figure}

\begin{figure}[!h]%
 \includegraphics[width=3in]{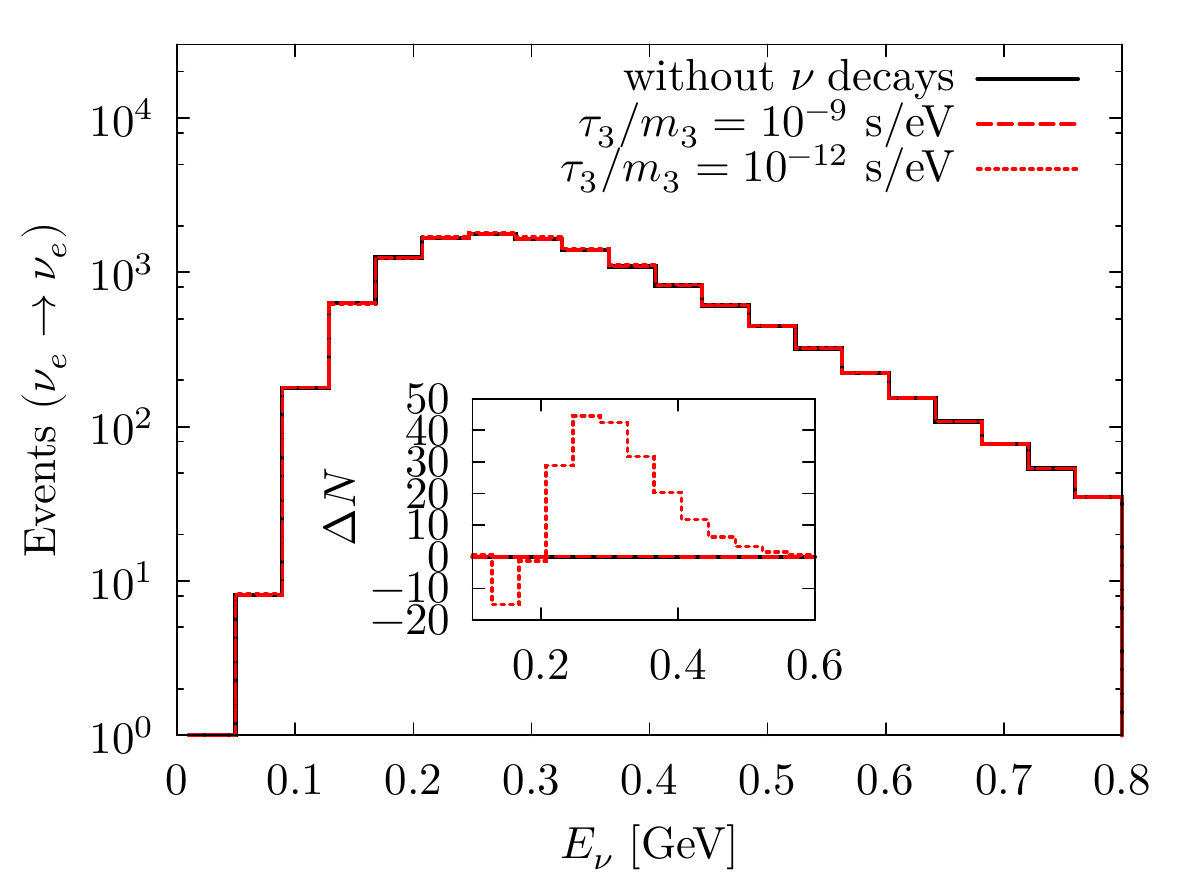}
  \includegraphics[width=3in]{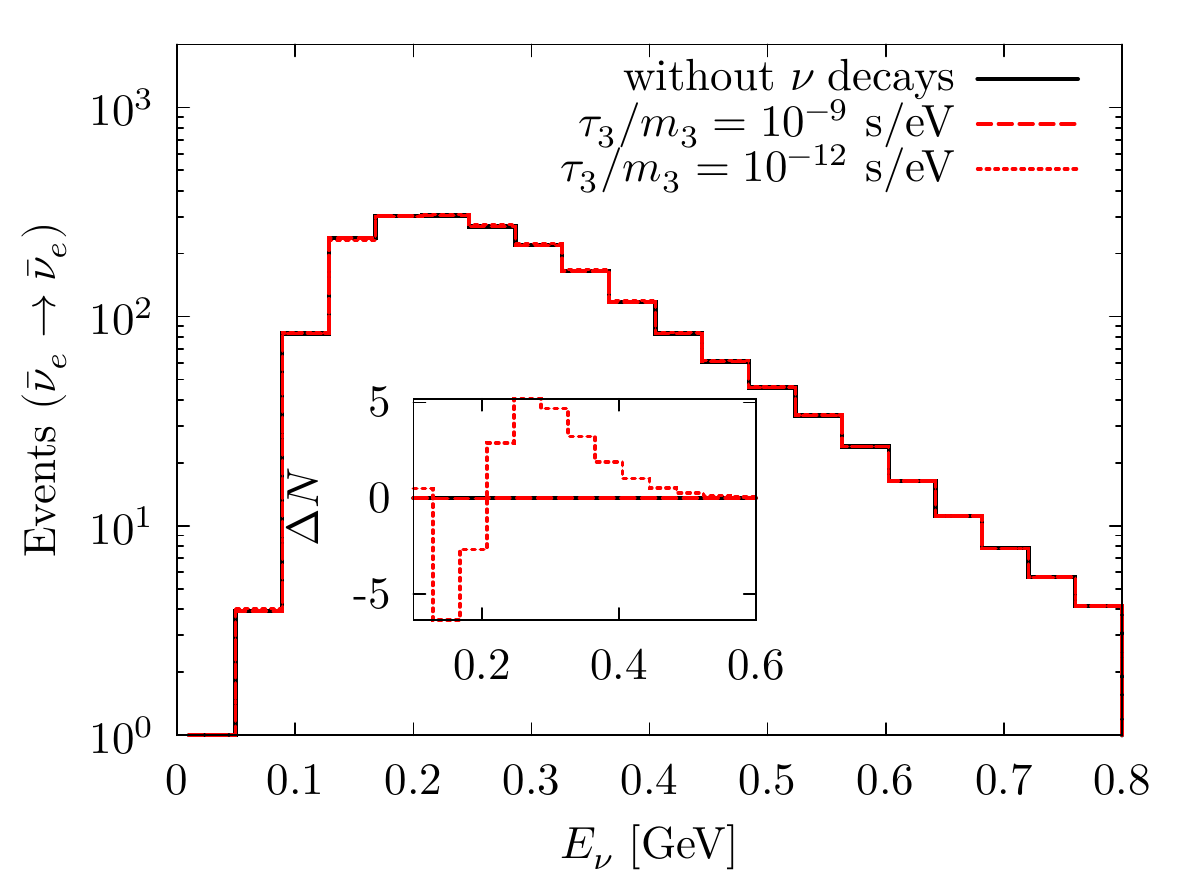}
 \includegraphics[width=3in]{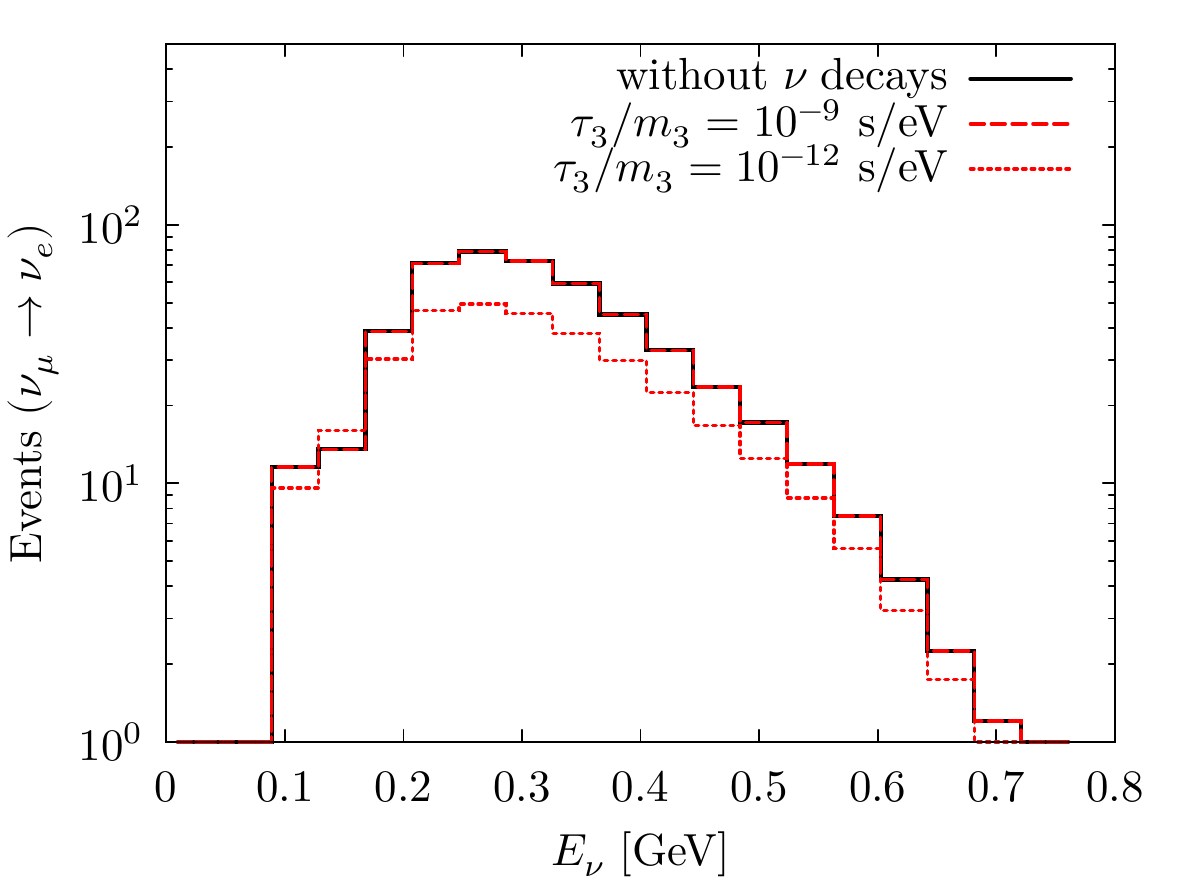}
  \includegraphics[width=3.in]{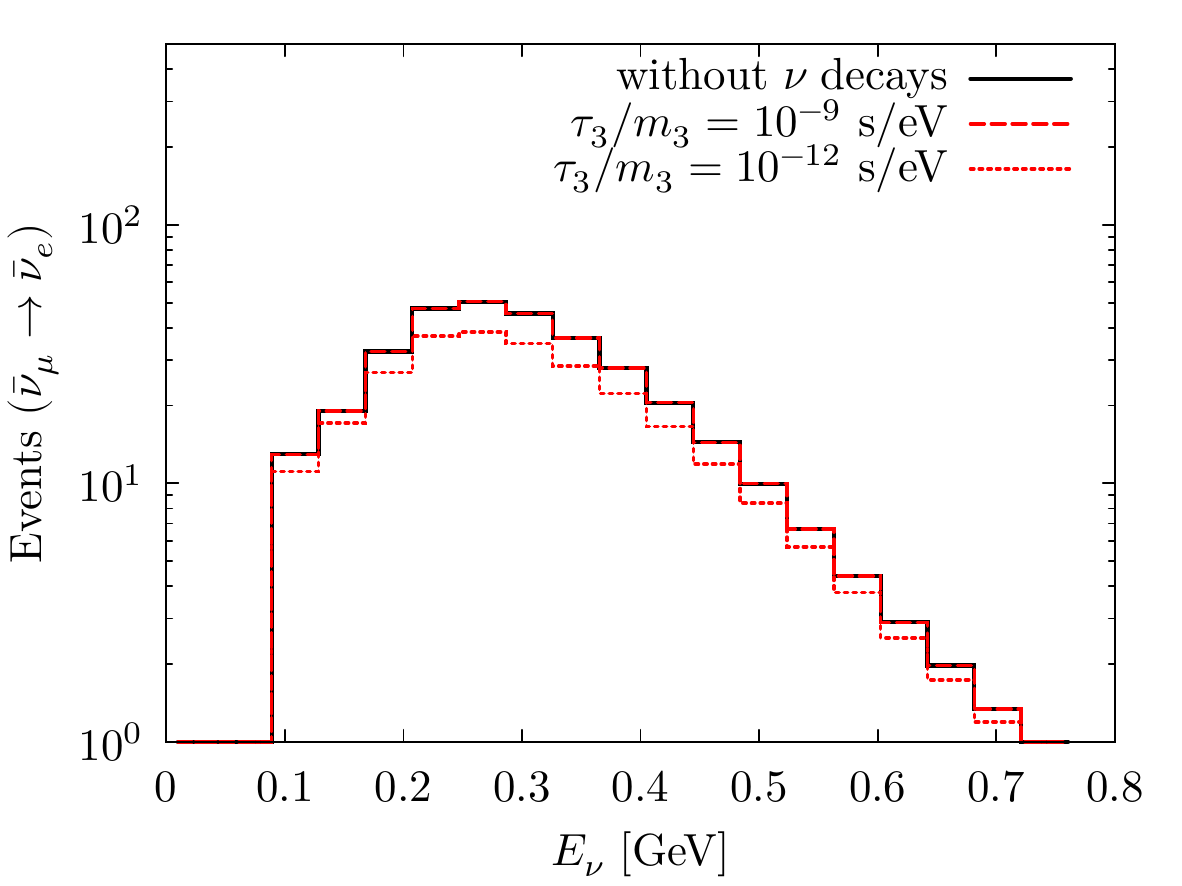}
 \caption{The spectra with $\tau_3/m_3=10^{-12}$~s/eV (red-dotted) and $=10^{-9}$~s/eV (red-dashed) and under the standard model (black) for MOMENT. Four channels are considered: $\nu_e\rightarrow\nu_e$ (upper-left), $\bar{\nu}_e\rightarrow\bar{\nu}_e$ (upper-right), $\nu_\mu\rightarrow\nu_e$ (right-left), and $\bar{\nu}_\mu\rightarrow\bar{\nu}_e$ (lower-right). The oscillation baseline is set at $150$ km. For clearness, in the result for the $\nu_e$ and $\bar{\nu}_e$ disappearance channels, we define $\Delta N$ as a difference between event rates with and without neutrino decays in each energy bin.}%
 \label{fig:event_e}
\end{figure}

We present the event spectra for each channels of MOMENT in Figs.~\ref{fig:event_mu} and \ref{fig:event_e}. Similar to Figs.~\ref{fig:prob} and \ref{fig:prob_bar}, the spectra for the case with $\tau_3/m_3=10^{-9}$ s/eV exactly overlap the spectra for the case without neutrino decays. The extreme case $\tau_3/m_3=10^{-12}$ s/eV is far from the black spectra, which are predicted assuming stable neutrinos. In the following, we focus on a comparison of results given different assumptions. We observe the advantage of the lower energy events, as the larger deviations from the spectra for the case without neutrino decays appear in the lower-energy bins. 
Comparing all panels in Figs.~\ref{fig:event_mu} and \ref{fig:event_e}, we are reminded of the conclusion from Sec.~\ref{sec:prob} that the muon-flavour disappearance channels are the most important ones for the measurement of $\tau_3/m_3$, as the larger deviations from the black spectra are observed.
In $\nu_\mu$ and $\bar{\nu}_\mu$ disappearance channels shown in Fig.~\ref{fig:event_mu}, we see both suppression and damping effects. The event rate decreases all the way in energy because of neutrino decays. However, the degree of deficit becomes larger around the maximum, while it gets smaller at the minimum. 
The change in these $\nu_\mu$ and $\bar{\nu}_\mu$ disappearance channels can be a few hundred events per bin, and much larger than those in the other six channels, in which the deficit is a few tens of events per bin. 
The overall decrease is also seen in $\nu_e\rightarrow\nu_\mu$ and $\bar{\nu}_\mu\rightarrow\bar{\nu}_e$ channels.
However, the number of events decreases in the lower energy bin but increases in the higher energy bin because of neutrino decays in the $\nu_e$ and $\bar{\nu}_e$ disappearance channels.
%
%
We see a reduction of event rates in most energy bins in $\bar{\nu}_e\rightarrow\bar{\nu}_\mu$, $\nu_\mu\rightarrow\nu_e$, $\bar{\nu}_e\rightarrow\bar{\nu}_\mu$ and $\bar{\nu}_\mu\rightarrow\bar{\nu}_e$, as shown in the lower panels of Figs.~\ref{fig:event_mu} and \ref{fig:event_e}. 

To sum up, it is clear that when we turn on neutrino decays with $\tau_3/m_3=10^{-12}$ s/eV, a distinct difference between the cases with and without decays can be easily measured by MOMENT. 
Invisible decays can wash out the extreme of neutrino oscillations. Therefore, the focus on the maximum or minimum can help us to detect the effect of neutrino decays.
%
Furthermore, the differences in $\nu_\mu$ and $\bar{\nu}_\mu$ disappearance channels are larger than the other six channels. This implies that the $\nu_\mu\rightarrow\nu_\mu$ and $\bar{\nu}_\mu\rightarrow\bar{\nu}_\mu$ channels will play an important role in the analysis. This could affect the precision measurement of neutrino mixing parameters such as $\theta_{23}$ and $\Delta m_{31}^2$ which are mostly involved in these channels. As a result, the other channels could help with a clarification of this bias induced by neutrino decays.
We eventually come up with the conclusion that MOMENT is expected to have high-level sensitivities to the lifetime of $\nu_3$ since they have multiple channels, and this exactly demonstrates the advantage of Gd-doped water Cherenkov technology.


\section{Results}
\label{sec:results}
Based on simulated event spectra with/without neutrino decays, we investigate the precision measurement on $\tau_3/m_3$ of MOMENT, and compare it with the reach by the current experiments. We also study the expected exclusion level to the stable neutrino hypothesis ($\tau_3/m_3=\infty$) assuming various true values of $\tau_3/m_3$. We further present our results on the impact of statistical error, systematic uncertainty and energy resolution. Finally, we study the contours at $3\sigma$ on the $\theta_{23}-\tau_3/m_3$, $\Delta m^2_{31}-\tau_3/m_3$ and $\theta_{23}-\Delta m^2_{31}$ planes.

\subsection{Bound on the lifetime of $\nu_3$}
\label{sec:constrain}

\begin{figure}[!h]%
\centering
 \includegraphics[width=4in]{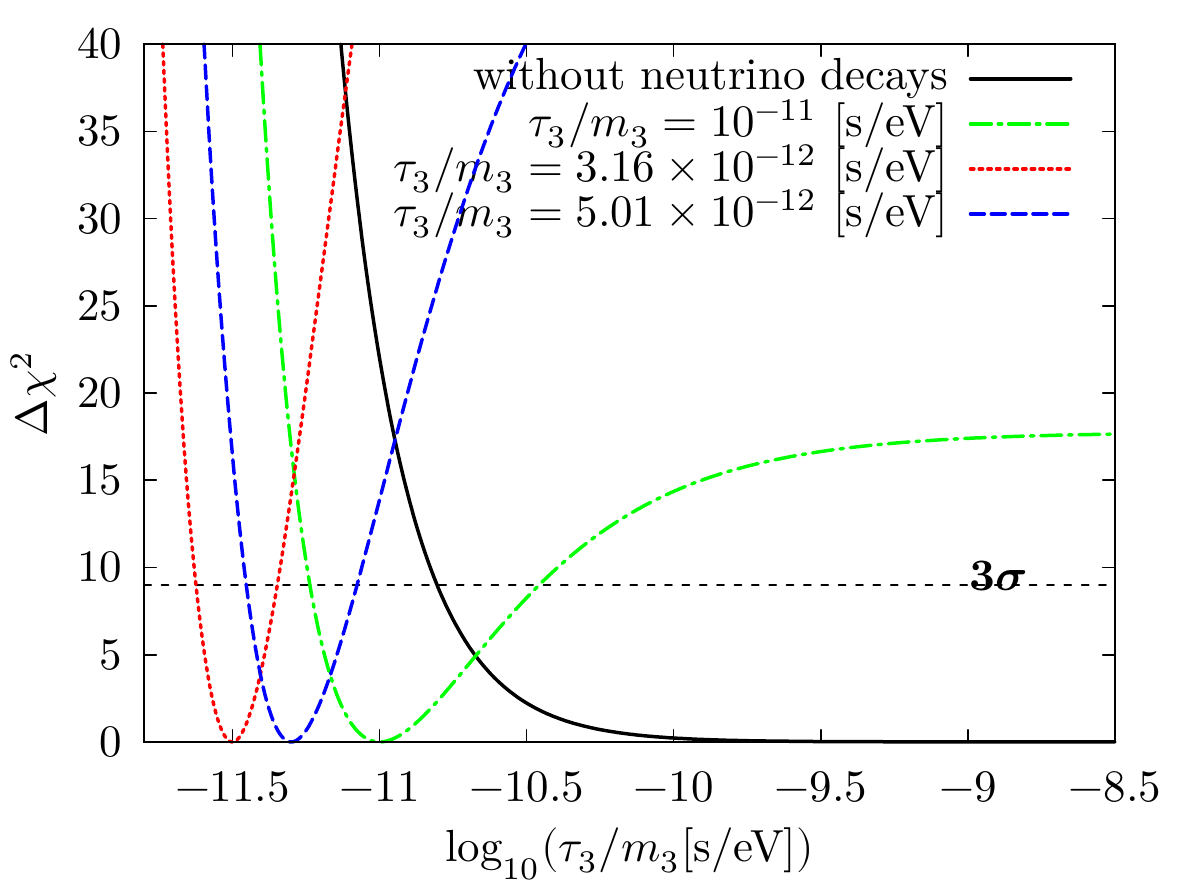}
  \caption{The $\Delta\chi^2=\chi^2-\chi^2_{min}$ as a function of the test value of $\tau_3/m_3$ where the input (true) value of $\tau_3/m_3$ is assumed to be $\infty$ (black solid), $10^{-11}$ (green dashed-dotted), $5.01\times10^{-12}$ (blue short-dashed) and $3.16\times10^{-12}~\text{s}/\text{eV}$ (red dotted) for MOMENT. We see that as the true value of $\tau_3/m_3$ gets smaller, the constraint becomes tighter, especially the upper bound.}%
 \label{fig:constrain_tau_m}
\end{figure}

In Fig.~\ref{fig:constrain_tau_m}, we show the constraint on $\tau_3/m_3$ for four different true values: $\tau_3/m_3=\infty$ (black solid), $10^{-11}$ (green dashed-dotted), $5.01\times10^{-12}$ (blue short-dashed), and $3.16\times10^{-12}~\text{s}/\text{eV}$ (red dotted). 
The latter three values are the current results from NOvA, T2K and the combined analysis of these two.
It is obvious that for larger neutrino-decay effects, the constraint becomes tighter. The appearance of the upper bound at $3\sigma$, which does not show up in the current measurements, is notable.
In the case of $\tau_3/m_3=10^{-11}$ s/eV, the lower (upper) bound at $3\sigma$ is at $\log_{10}(\tau_3/m_3)\sim -11.25$ ($-10.5$).
With $\tau_3/m_3=5.01\times10^{-12}~\text{s}/\text{eV}$, the $3\sigma$ constraint is about $10^{-11.5}-10^{-11.1}~\text{s}/\text{eV}$, while with $\tau_3/m_3=3.16\times10^{-12}~\text{s}/\text{eV}$ the $3\sigma$ uncertainty runs from $\sim10^{-11.65}$ to $\sim10^{-11.35}~\text{s}/\text{eV}$. 
{\color{black}
%
The whole behaviour of $\Delta\chi^2$ is that starting from the true value, it climbs to infinity when $\tau_3/m_3$ gets smaller, while $\Delta\chi^2$ approaches to a certain value when $\tau_3/m_3\rightarrow\infty$. The behaviour can be understood in Figs.~\ref{fig:event_mu} and \ref{fig:event_e}. When $\tau_3/m_3$ is larger enough, the spectra behave the same as those for the stable-neutrino case. Therefore, $\Delta\chi^2$ approaches to a certain value when $\tau_3/m_3\rightarrow\infty$. We note that the behaviour of $\Delta\chi^2$ looks symmetric for $\tau_3/m_3=3.16\times10^{-12}$ [s/eV] in Fig.~\ref{fig:constrain_tau_m}, but does not for the larger value of $\tau_3/m_3$. It is because in the case with $\tau_3/m_3=3.16\times10^{-12}$ [s/eV] $\Delta\chi^2$ is approaching to $\sim120$ when $\tau_3/m_3\rightarrow\infty$. The range of $\Delta\chi^2$ shown in Fig.~\ref{fig:constrain_tau_m} is near the bottom. Therefore, the behaviour of $\Delta\chi^2$ looks symmetric for $\tau_3/m_3=3.16\times10^{-12}$ [s/eV].}
%

\begin{figure}[!h]%
 \includegraphics[width=3in]{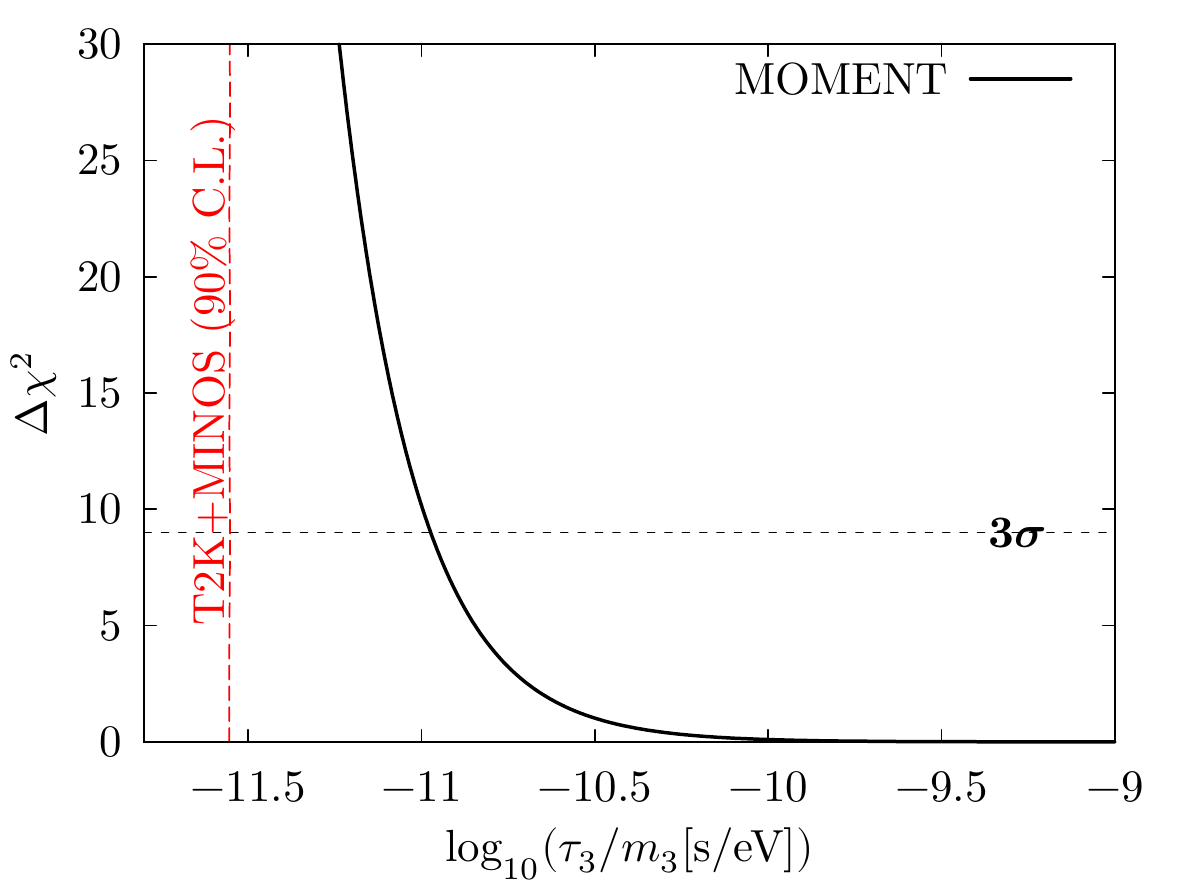}
 \includegraphics[width=3in]{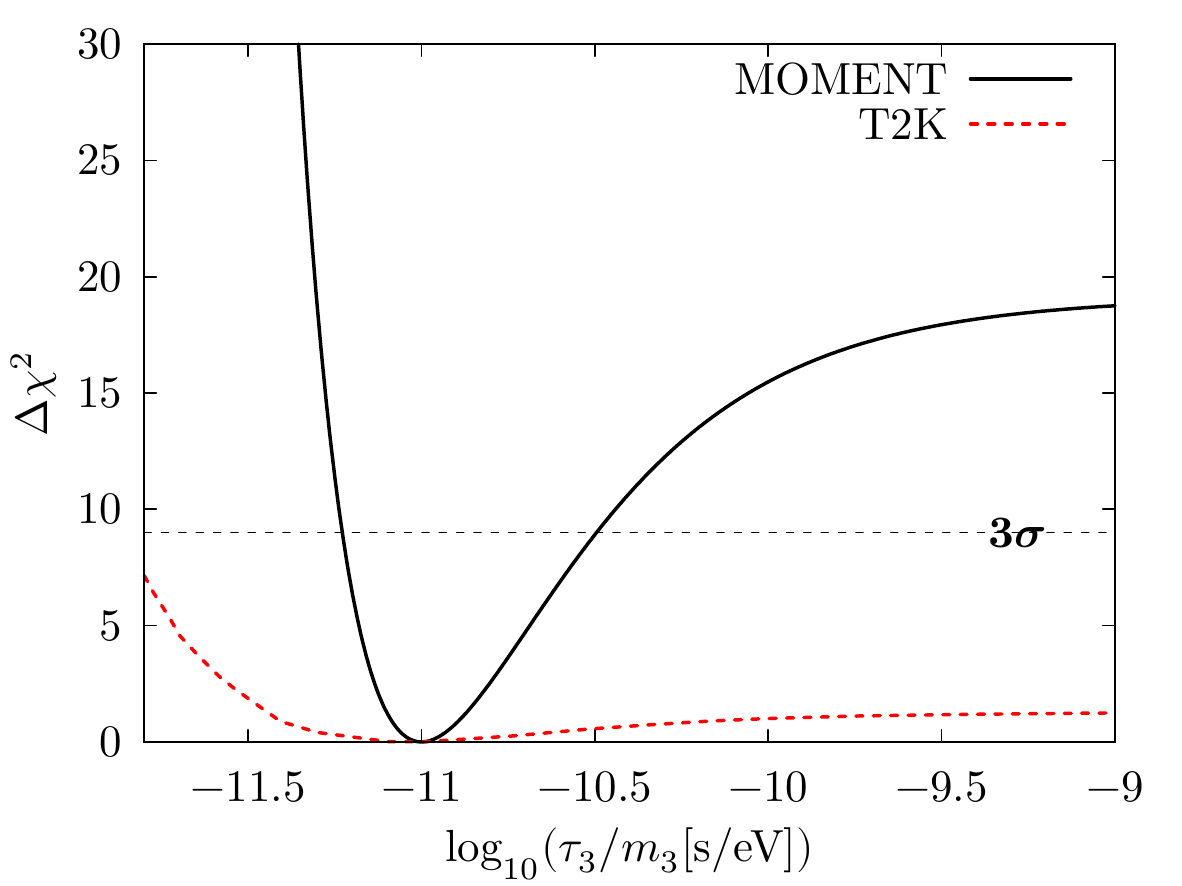}
  \includegraphics[width=3in]{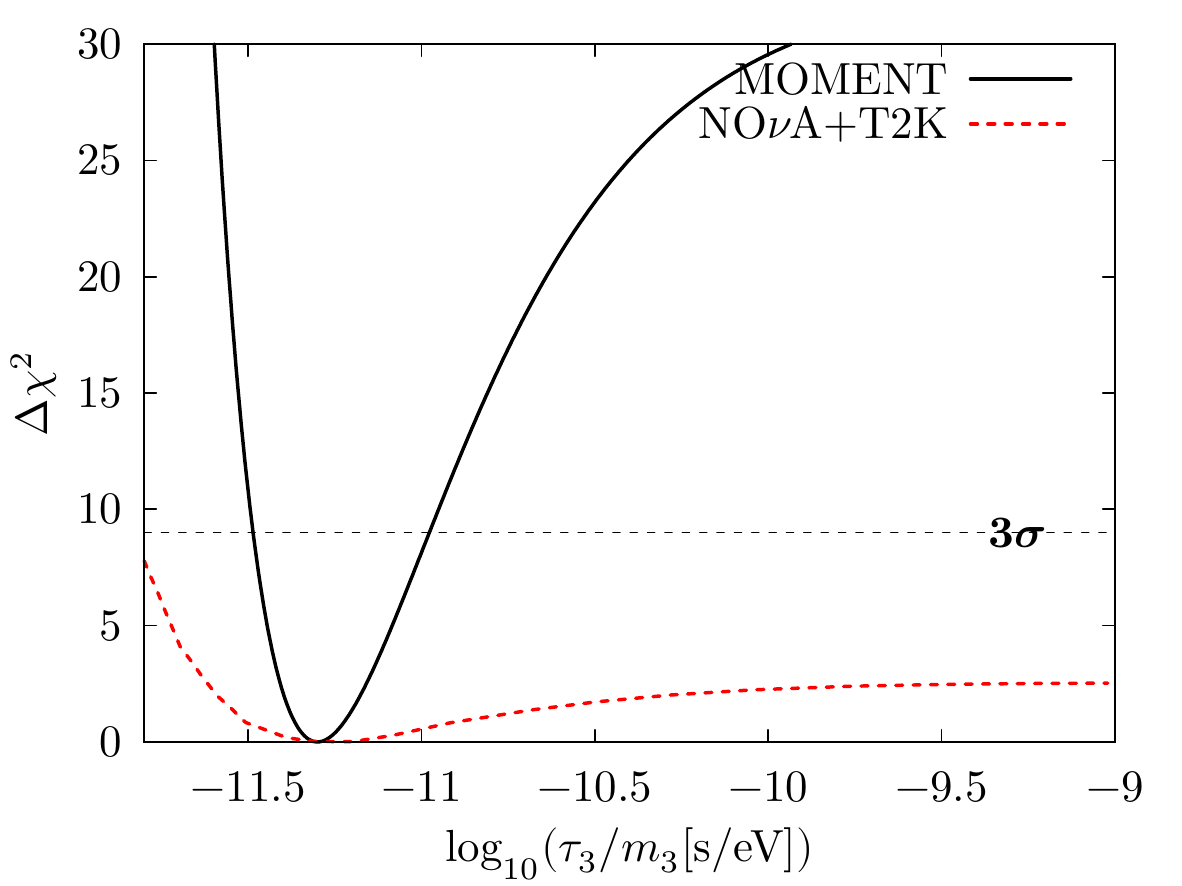}
  \includegraphics[width=3in]{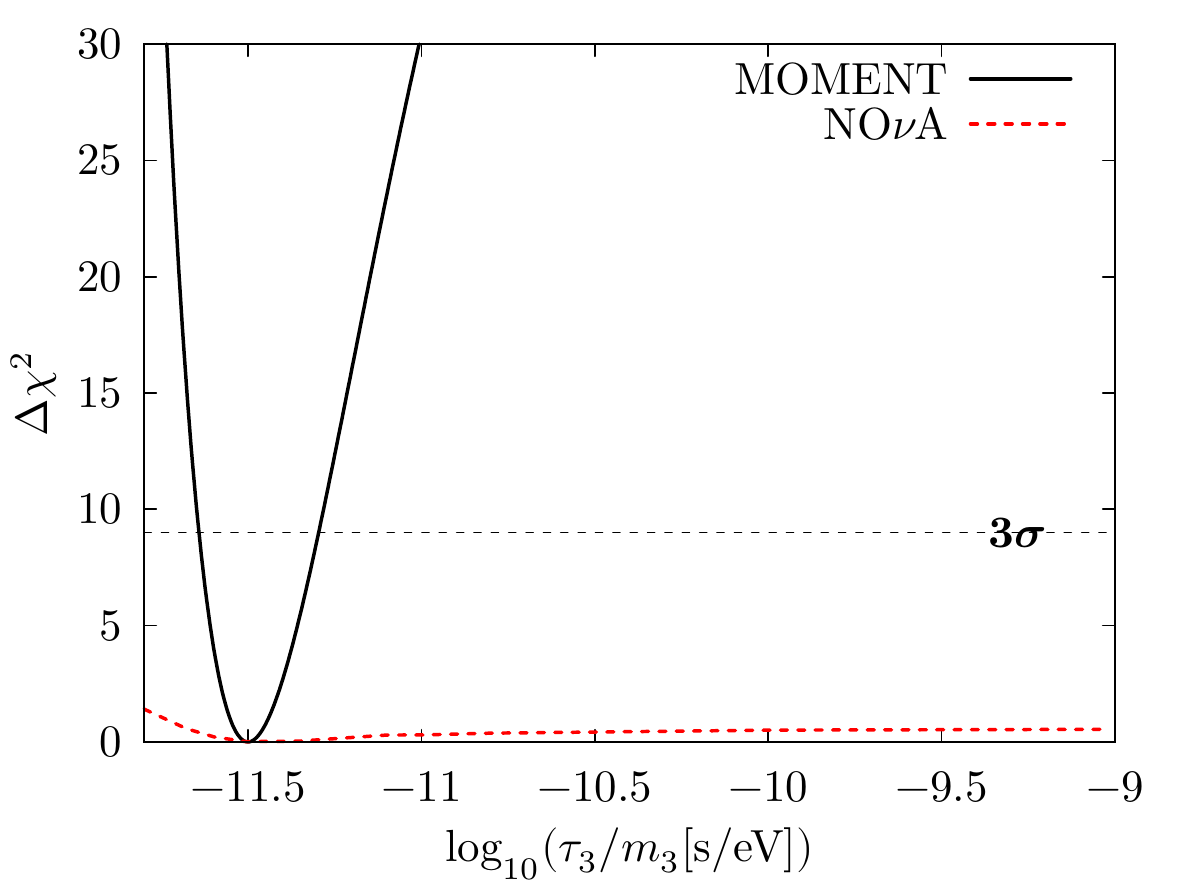}
  \caption{The $\Delta\chi^2=\chi^2-\chi^2_{min}$ as a function of the test value of $\tau_3/m_3$ where the inputed true value of $\tau_3/m_3$ is assumed to be $\infty$ (upper left), $10^{-11}$ (upper right), $5.01\times10^{-12}$ (lower left) and $3.16\times10^{-12}~\text{s}/\text{eV}$. The black solid curve is for MOMENT. The short-dashed curves, taken from Ref.~\cite{Choubey:2018cfz}, correspond to current experiments: the upper-left and upper-right panels are the combination of T2K and MINOS and T2K, respectively, while the lower two panel are the combination of T2K and NOvA (left) and NOvA (right).}%
 \label{fig:compare_tau_m}
\end{figure}

In Fig.~\ref{fig:compare_tau_m} we compare the result from MOMENT (black curve) with the current experiments (red short-dashed curves), which are taken from Ref.~\cite{Choubey:2018cfz}.
The upper-left panel shows the constraint assuming the case without neutrino decays. As we can see, the bound at $3\sigma$ for $\tau_3/m_3$ is pushed up by about one order of magnitude  from the bound at $90\%$ C.L. for the combination of T2K and MINOS. 
Except for the upper left panel, the difference from the curves in Fig.~\ref{fig:constrain_tau_m} is that we use the same true values for $\theta_{23}$ and $\Delta m_{31}^2$ as the best fit of Ref.~\cite{Choubey:2018cfz} in this figure. 
%
%
The most striking feature of MOMENT we see in this figure is that it provides the upper bound for $\tau_3/m_3$ measurement at $3\sigma$, while the lower bound is also greatly reduced. In the other words, instead of giving us a lower bound, MOMENT provides a complete range with the upper and lower limits at a considerable confidence level. The upper bound is important for excluding the case without neutrino decays, if  the neutrino decay is confirmed.

\begin{figure}[!h]%
\centering
 \includegraphics[width=4in]{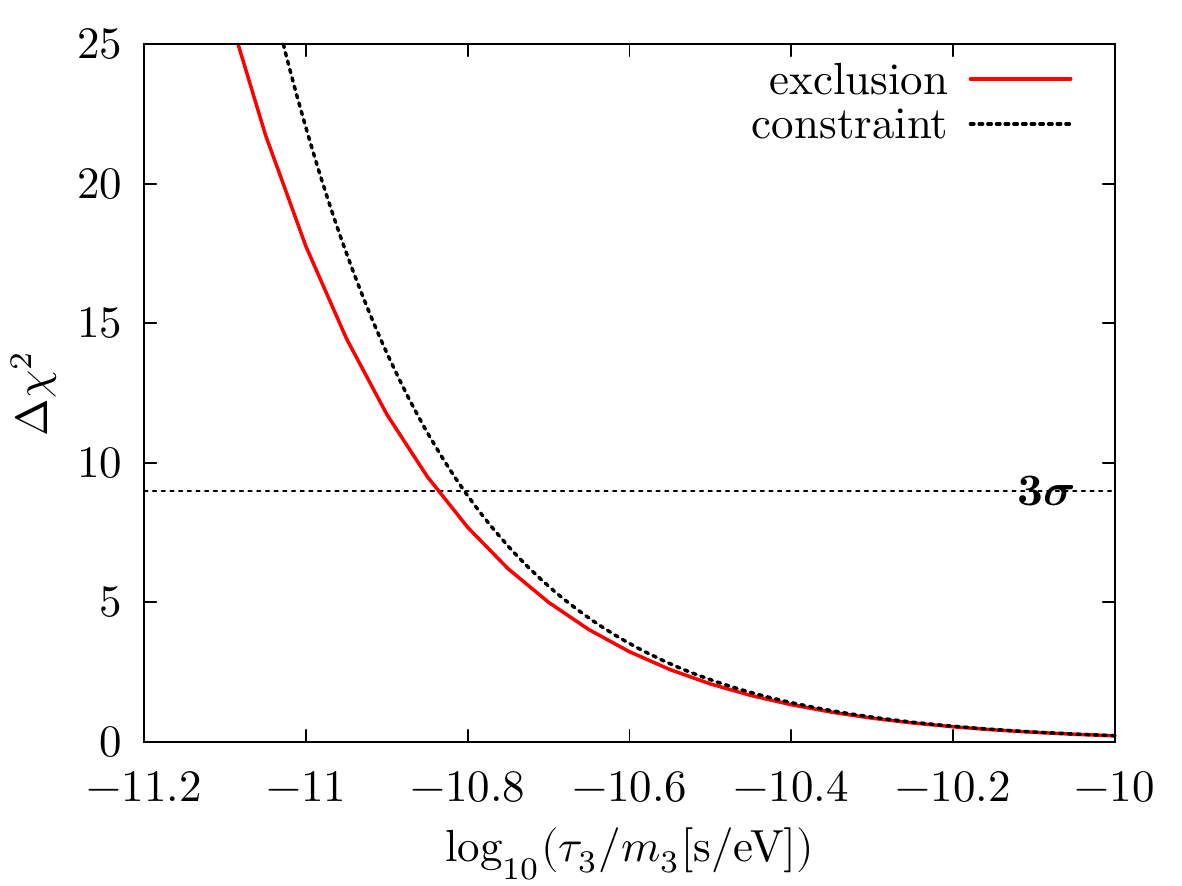}
  \caption{The exclusion ability of the stable-neutrino case $m_3/\tau_3=0$ (the red curve) and the constraint on $\tau_3/m_3$ assuming the stable-neutrino case (the black curve). The solid curve corresponds to MOMENT. The exclusion ability is defined as $\Delta\chi^2$ for the hypothesis $m_3/\tau_3=0$ for various true values (the x-axis value). We find the exclusion ability could reach $3\sigma$ while $\tau_3/m_3$ is below $10^{-10.8}$~s/eV for MOMENT.}%
 \label{fig:ex0}
\end{figure}

From Fig.~\ref{fig:constrain_tau_m}, it is natural to expect that these experiments have a great ability to exclude the stable-neutrino hypothesis $\tau_3/m_3=\infty$. We therefore discuss while the true $\tau_3/m_3$ is not infinity, how much MOMENT can exclude the stable-neutrino hypothesis, and therefore find a \textit{hint} of new physics.\footnote{We call the tension between the experimental result and the stable-neutrino prediction ``\textit{hint}''.} We show our results in Fig.~\ref{fig:ex0}, in which the red curve is the exclusion ability for MOMENT. The statistical quantity we are studying is $\Delta\chi^2$ for the hypothesis $m_3/\tau_3=0$ assuming the various true values $\tau_3/m_3$ (x-axis). We also compare these results with the constraint on $\tau_3/m_3$ assuming the case with neutrino decays (black curves). 
We find that if in the nature $\log_{10}({\tau_3/m_3}[\text{s/eV}])\sim-10.85$, MOMENT can detect a ``\textit{hint}'' at around $3\sigma$. These $\tau_3/m_3$ values are larger than our current discovery from T2K and NOvA. This means MOMENT could be sensitive enough to claim a ``\textit{hint}'' if the current results are confirmed.

\subsection{Impact of the total running time, systematic uncertainty, and energy resolution}\label{sec:exp_impacts}

\begin{figure}[!h]%
\centering
 \includegraphics[width=4in]{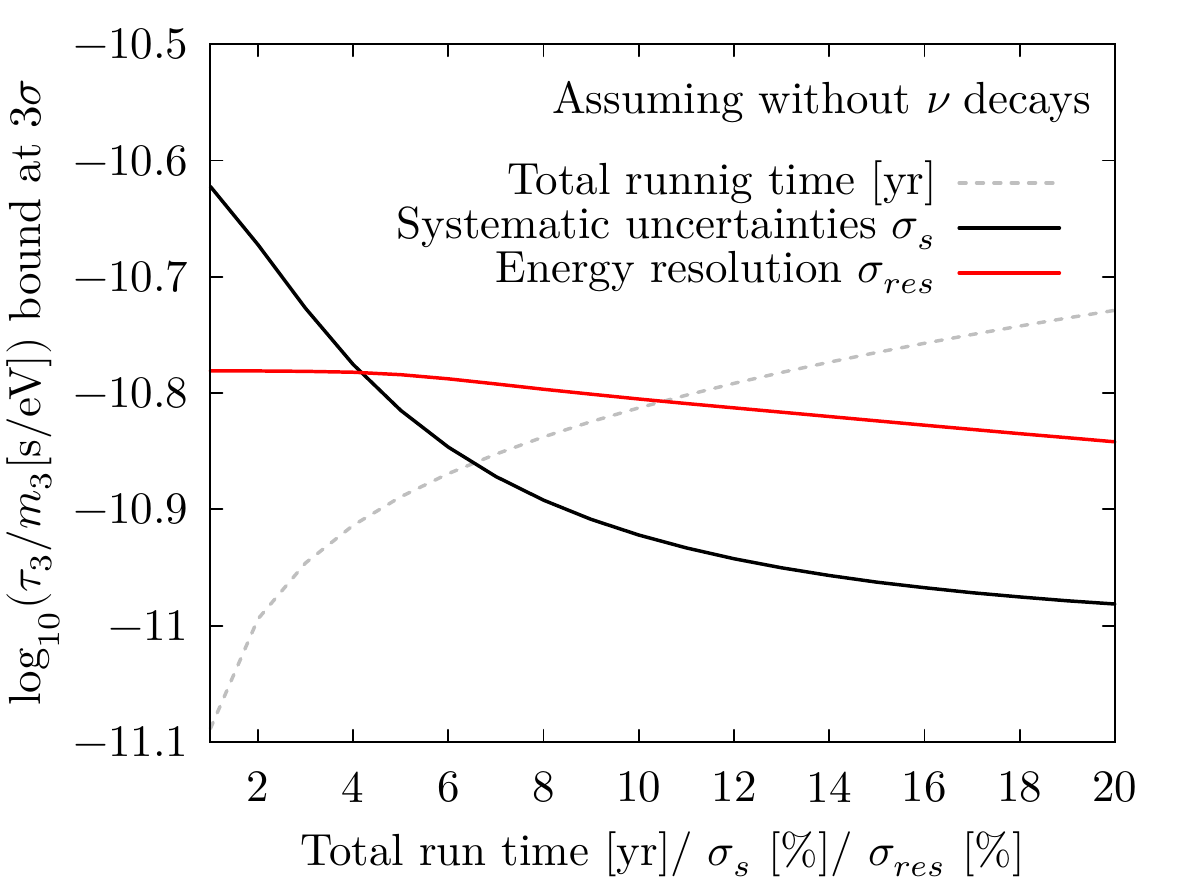}
  \caption{The constraint at $3\sigma$ on $\tau_3/m_3$ assuming the stable-neutrino case against the total running time (the short-dashed grey curve), the size of systematic uncertainty $\sigma_s$ (black), and the energy resolution $\sigma_{res}$ (red). We focus on the total running time from 1 to 20 years, while $\sigma_s$ and $\sigma_{res}$ vary in the range $[1\%, 20\%]$.}%
 \label{fig:exp_impacts}
\end{figure}


\begin{figure}[!t]%

 \includegraphics[width=3in]{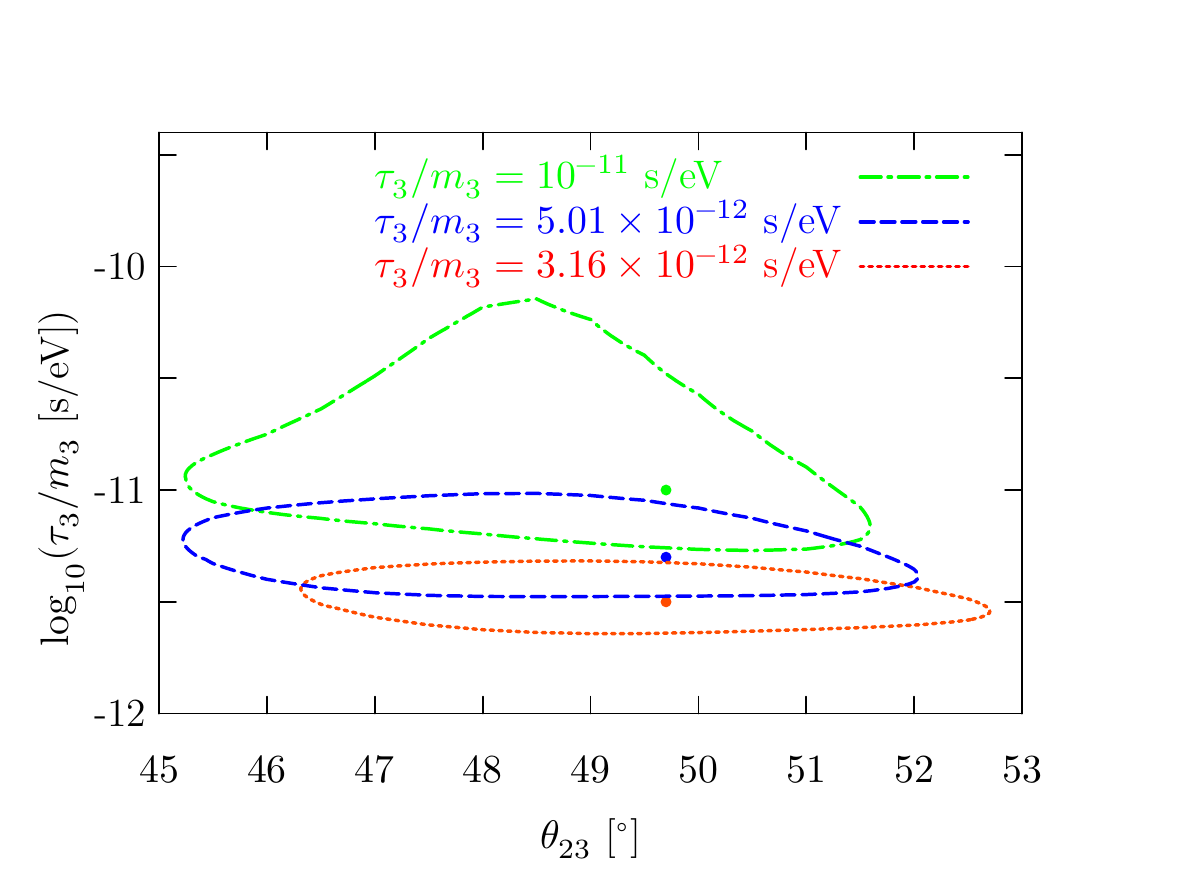}
  \includegraphics[width=3in]{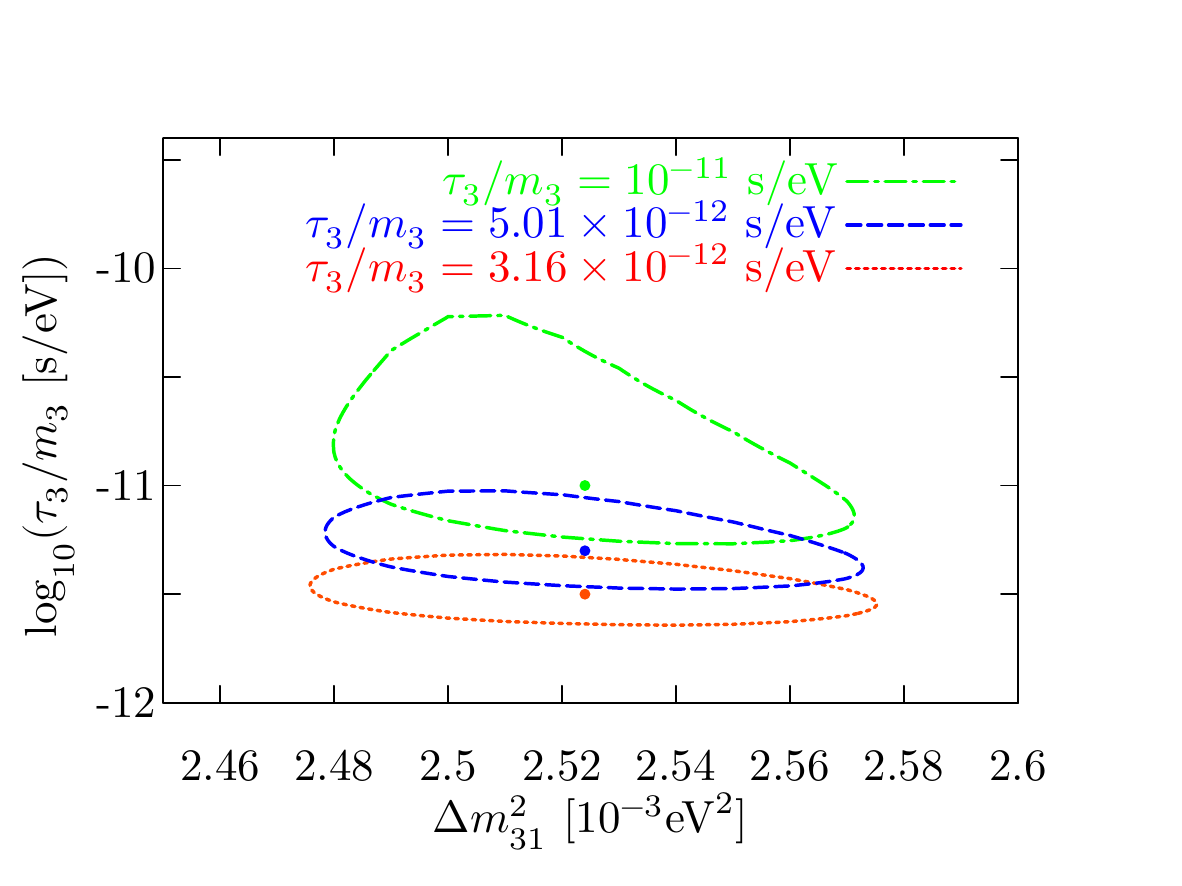}\\
 \includegraphics[width=3in]{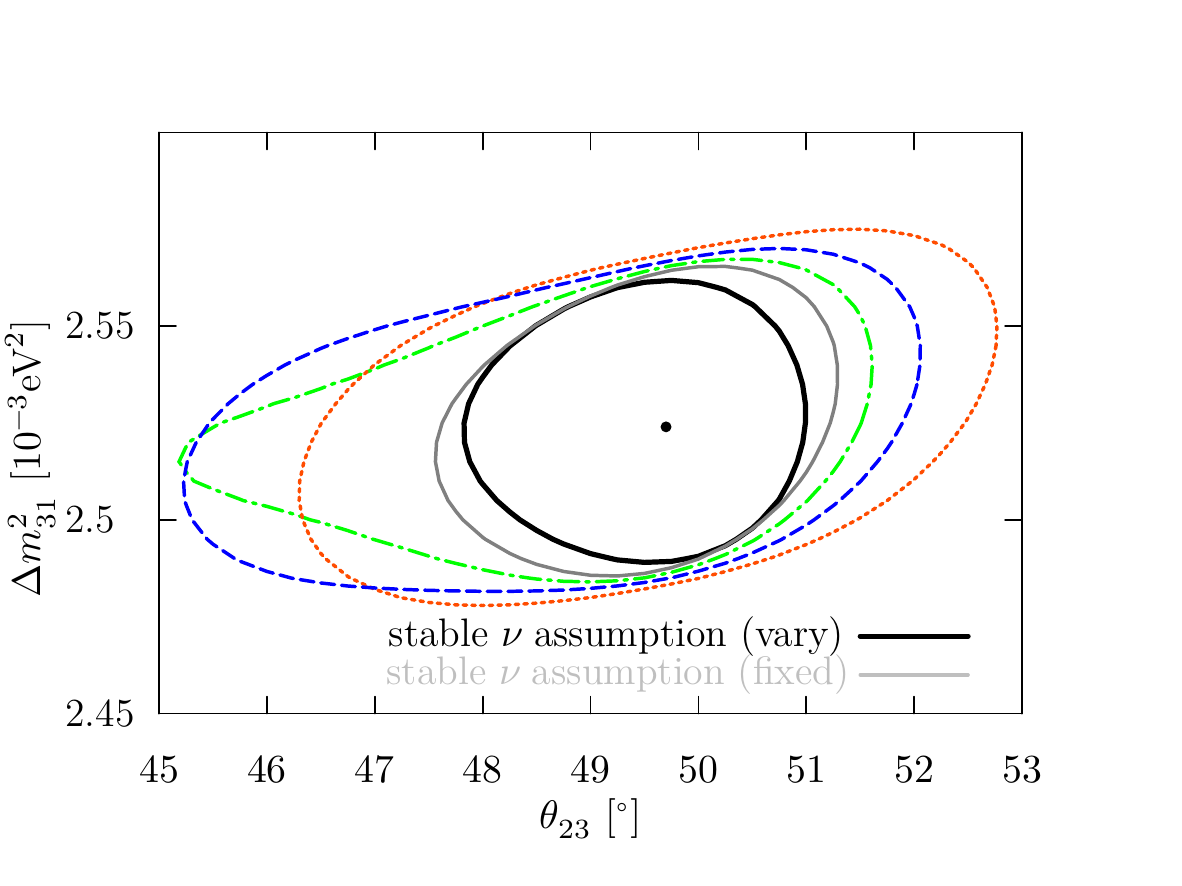}$~~~~~~~~~~~~~$\\
  \caption{The exclusion contour at $3\sigma$ on the planes any two of $\log_{10}(\tau_3/m_3[\text{s/eV}])$ and $\theta_{23}$ (left) and $\Delta m^2_{31}$ (right). We show for different true values: $\tau_3/m_3=10^{-11}$ (dashed-dotted green), $5.01\times10^{-12}$ (short-dashed blue) and $3.16\times10^{-12}$ (dotted red) s/eV. In the lower panel, we further consider two scenarios --- $\tau_3/m_3$ fixed at $\infty$ (black) and let this parameter vary (grey).}%
 \label{fig:th23_ldm_tau}
\end{figure}

We are interested in studying the impact of the total running time (the short-dashed grey curve), the systematic uncertainty (the black solid curve) and the energy resolution (the red solid curves) in Fig.~\ref{fig:exp_impacts}.
We present the constraint power assuming the case without neutrino decays at the $3\sigma$ confidence level.\footnote{We also undergo the same study for the $3\sigma$ exclusion ability to the stable-neutrino hypothesis. The results are almost the same as those shown in Fig.~\ref{fig:exp_impacts}.} Going through the total running time ($\nu$-mode $+$ $\bar{\nu}$-mode) from $1$ to $20$ years, a $3\sigma$ bound can improve from about $\tau_3/m_3=10^{-11}$ to $10^{-10.7}$ s/eV. It soars from $\log_{10}(\tau_3/m_3 [\text{s/eV}])=-11$ to about $\sim-10.9$ at the fourth year before a slow climb to $-10.7$ at the twentieth year. This means that once it runs for more than $4$ years, it gets more difficult to improve the sensitivity by increasing the running time.
Moving to the impact of systematic uncertainties, we vary the size of the normalisation uncertainty\footnote{The systematic uncertainty, in which we are interested, is the combination of that of fiducial detector volume, flux error for signals, and so on.} $\sigma_s$ from $1$ to $20\%$ for all channels. By decreasing $\sigma_s$, we can improve the $3\sigma$ bound from $\tau_3/m_3=10^{-11}$ to $10^{-10.6}$ s/eV.  The improvement rises quickly when $\sigma_s<5\%$ --- from $\log_{10}(\tau_3/m_3[\text{s}/\text{eV}])\sim-10.85$ for $\sigma_s=5\%$ to  $-10.6$ for $1\%$.
Finally, we see relatively small impacts by improving the energy resolution.

We find an important result by comparing two curves, representing the impact of the total running time and $\sigma_{s}$. Our default setting for MOMENT is the case with $10$ years for the total running time and roughly the point for $\sigma_s=5\%$; comparing to two curves, we can see improving $\sigma_s=1\%$ can improve better ($\log_{10}(\tau_3/m_3 [\text{s/eV}])=-10.6$) than that by doubling the total running time ($\log_{10}(\tau_3/m_3 [\text{s/eV}])=-10.7$). Then, we further conclude that improving our understanding of systematic uncertainties is more important than doubling the total running time.

\subsection{Precision measurements of $\tau_3/m_3$ with $\theta_{23}$ and $\Delta m^2_{31}$}

As we see in Fig.~\ref{fig:prob}, the measurement of $\tau_3/m_3$ largely depends on the disappearance channel, which is sensitive to $\theta_{23}$ and $\Delta m_{31}^2$. We are therefore interested in the performance of $3\sigma$ contours on the $\tau_3/m_3-\theta_{23}$ (upper-left), $\tau_3/m_3-\Delta m^2_{31}$ (upper-right) and $\theta_{23}-\Delta m^2_{31}$ (lower) planes in Fig.~\ref{fig:th23_ldm_tau}. We assume three true values: $\tau_3/m_3=10^{-11}$ (dashed-dotted green), $5.01\times10^{-12}$ (short-dashed blue) and $3.16\times10^{-12}$ (red dotted)~s/eV. Thanks to the high precision of the $\tau_3/m_3$ measurement, we see a complete contour, instead of a band as what we see in current fitting result, shown in Ref.~\cite{Choubey:2018cfz}. On average, the precision at $3\sigma$ of $\theta_{23}$ is almost $3-3.5^\circ$ for MOMENT. We observe some impact from the true $\tau_3/m_3$ value on the $\theta_{23}$ measurement. The $3\sigma$ uncertainty of $\Delta m^2_{31}$ is about $0.05\times 10^{-3}$eV$^2$.
%
%
We further study the $3\sigma$ contour on the $\theta_{23}-\Delta m^2_{31}$ plane (the lower panel). We also include results for the stable neutrino case. We consider two scenarios --- $\tau_3/m_3$ fixed at $\infty$ (black) and let this parameter vary (grey). It is obvious that the impact of neutrino decays mainly worsens the measurement of $\theta_{23}$ from $\sim1.5^\circ$ to $\sim 3-3.5^\circ$. In comparison, there is little impact on the measurement of $\Delta m^2_{31}$. We also see a little correlation, once we include $\tau_3/m_3$ into fitting.

\section{Summary}
\label{sec:summary}
In this paper we have considered the third neutrino mass eigenstate $\nu_3$ decaying to invisible states in MOMENT, using eight channels of neutrino oscillation ($\nu_e\rightarrow \nu_e$, $\nu_e\rightarrow \nu_{\mu}$, $\nu_{\mu} \rightarrow \nu_e$, $\nu_{\mu} \rightarrow \nu_{\mu}$ and their CP-conjugate partners) with the help of the following detection processes in a Gd-doped Cherenkov detector: $\nu_e + n \rightarrow p + e^-$, $\bar{\nu}_{\mu} + p \rightarrow n + \mu^+$, $\bar{\nu}_e + p \rightarrow n + e^+$, and $\nu_{\mu} + n \rightarrow p + \mu^- $. 
Neutrino decays cause suppression and damping effects on neutrino oscillation probabilities, and could be measured in the reconstructed energy spectra of MOMENT, especially in $\nu_\mu$ and $\bar{\nu}_\mu$ disappearance channels. And we have found that focusing on the maximum or minimum is a strategy to measure these effects. 
Events with lower neutrino energy do not only avoid the sizeable matter effect, but also enhance the effects caused by neutrino decays. 
We have simulated the MOMENT experiment and found outstanding potential to constrain the $\tau_3/m_3$ parameter in Fig.~\ref{fig:constrain_tau_m}. Given the best-fit values hinted by T2K and NOvA~\cite{Choubey:2018cfz}, we have found that MOMENT would improve the precision measurement of invisible neutrino decays. We reach an interesting conclusion that if the current best fit discovered in~\cite{Choubey:2018cfz} is confirmed, the standard non-decay scenario can be excluded with a statistics level higher than $3\sigma$. 
At $3\sigma$ confidence level, the projections of $\theta_{23}-\log_{10}(\tau_3/m_3)$, $\Delta m_{31}^2-\log_{10}(\tau_3/m_3)$ and $\theta_{23}-\Delta m^2_{31}$ have demonstrated little correlations between $\theta_{23}$ and $\Delta m^2_{31}$.
The impact of neutrino decays mainly decrease the $3\sigma$ precision of $\theta_{23}$ by $1-1.5^\circ$.
%

We have further investigated the impact of statistical and systematic uncertainties by varying the total running time, changing the size of the normalisation uncertainty $\sigma_s$ and energy resolution respectively. We have demonstrated the $3\sigma$ constraint assuming the standard non-decay scenario. By increasing the total running time or reducing the systematic uncertainties, we will improve the sensitivity in invisible neutrino decays. A comparison of two methods has guided us to the conclusion that reducing systematic uncertainties is more important than increasing the total running time in the MOMENT experiment. We have also checked that there is no sizeable impact from improved energy resolution in the detector.

As MOMENT has outstanding potential to measure neutrino decays, we also have to emphasize that future atmospheric and astrophysical neutrino experiments will significantly improve the current understanding of neutrino decays. They are complementary to each other, though.

\acknowledgments
This work is supported in part by the National Natural Science Foundation of China under Grant No. 11505301 and No. 11881240247. JT appreciates ICTP's hospitality and scientific activities during the workshop PANE2018. We would like to thank Thomas Hahn for communications and providing a package to diagonize non-hermitain matrices. 
We would like to thank the accelerator working group of MOMENT for useful discussions and for kindly providing flux files for the MOMENT experiment. Finally, we appreciate Dr.~Neil Drouard Raper's help to improve the readability of our paper.



\end{document}